\documentclass{aastex}
\usepackage{spr-astr-addons}
\usepackage{url}
\usepackage{hyperref}
\hypersetup{colorlinks,linkcolor={blue},citecolor={blue},urlcolor={red}}

\RequirePackage{color}

\begin{document}

\title{Resistive Hot Accretion Flows with Anisotropic Pressure}
\slugcomment{Not to appear in Nonlearned J., 45.}
\shorttitle{Resistive Hot Accretion Flows}
\shortauthors{Ghoreyshi and Khesali}

\author{S. M.  Ghoreyshi} \and \author{A. R. Khesali}
\email{\emaila}{smghoreyshi64@gmail.com}
\affil{Department of Physics, University of Mazandaran, Babolsar,
Iran.}
%
%


\begin{abstract}

Since the collisional mean free path of charged particles in hot accretion
flows can be significantly larger than the typical length-scale of the
accretion flows, the gas pressure is anisotropic to magnetic field lines.
For such a large collisional mean free path, the resistive dissipation
can also play a key role in hot accretion flows. In this paper, we study
the dynamics of resistive hot accretion flows in the presence of anisotropic
pressure. We present a set of self-similar solutions where the flow variables
are assumed to be a function only of radius. Our results show that the
radial and rotational velocities and the sound speed increase considerably
with the strength of anisotropic pressure. The increase in infall velocity
and in sound speed are more significant if the resistive dissipation is
taken into account. We find that such changes depend on the field strength.
Our results indicate that the resistive heating is $10\%$ of the heating
by the work done by anisotropic pressure when the strength of anisotropic
pressure is 0.1. This value becomes higher when the strength of anisotropic
pressure reduces. The increase in disk temperature can lead to heating
and acceleration of the electrons in such flows. It helps us to explain
the origin of phenomena such as the flares in Galactic Center Sgr A*.
\end{abstract}


\section{Introduction}

In standard accretion disks \citep{Shakura1973}, advection is not considered
and the energy released due to dissipative process is radiated away. The
standard disks are relatively cold and cannot explain some phenomena in
systems such as Galactic Center Sgr A* \citep{Kato2008}. Therefore, other
models such as the radiatively inefficient accretion flow (RIAF) model
have been introduced \citep{Narayan1994,Kato2008}. In such disks, the released
energy is retained within the accreting flows and is advected radially
inward. This leads to hot accretion flows and produces high-energy emission.
In addition to extremely low-luminosity active galactic nuclei (LLAGNs)
such as Sgr A* \citep{Yuan2002,Yuan2014} and M 87 \citep{Park2019}, hot
accretion flows can be an appropriate model to the quiescent and hard states
of black hole X-ray binaries
\citep{Done2007,Qiao2009,Zhang2012,Yan2017}.

The numerical simulations have revealed that strong outflows can be launched
in hot accretion flows
\citep{Ohsuga2009,Yuan2012a,Yuan2012,Yuan2015,Bu2016a,Bu2016b}. The presence
of outflows has also been confirmed by the observations of LLAGNs and X-ray
binaries
\citep{Tombesi2010,Crenshaw2012,Wang2013,Tombesi2014,Cheung2016,Homan2016,Ma2019,Park2019}.
The outflows can extract mass, angular momentum and energy from the disk
\citep{Pudritz1985,Knigge1999,Xue2005,Cao2016,Cao2019} and strongly affect
the structure of accreting flows. For instance, the disk gas surrounding
the black hole can be pushed away by the outflows. This extraction causes
the accretion rate of the black hole to decrease
\citep{Yuan2018,Bu2019ApJ}. Several attempts to explore the effects of
outflow on the dynamics of hot accretion flows have been made over the
last several years. The results of previous works indicate that the outflows
can be an effective cooling agent in hot accretion flows
\citep[e.g.,][]{Bu2009,Mosallanezhad2013,Bu2019Uni,Ghoreyshi2020}.

Recent findings indicate that there is an orbiting hot spot nearby
the event horizon of Sgr A* \citep{Gravity2018}. Non-thermal emission related to
such hot spots is believed to be due to magnetic reconnection
\citep{Ding2010,Ripperda2019mnras} and the ideal MHD is unable to explain
non-thermal emission. We require dissipative (non-ideal) mechanisms to
describe such observational phenomena in accretion systems. There are three
regimes in non-ideal MHD due to the relative drifts of different charged
species with respect to the neutral particles; ohmic dissipation, Hall
effect, and ambipolar diffusion. Although ohmic dissipation leading to
a decoupling of the magnetic field from all charged particles can be important
in the innermost parts of the disks around protostars, ambipolar diffusion
is the dominant non-ideal MHD effect in the outer regions of such disks
and causes the magnetic field to decouple from the neutral gas. But the
Hall effect occurs when the magnetic field decouples from ions and charged
dust grains. This non-ideal effect, which is an intermediate regime between
ambipolar diffusion and ohmic resistivity, plays an important role at other
parts of such disks \citep[see][]{Braiding2012}.

Flows of disks around dwarf novae
\citep{Fleming2000,Sano2002,Scepi2018} and protoplanetary disks
\citep{Bethune2017,Mori2019,Wang2019,Das2021} which are cold and partially
ionized can be affected by all of these non-ideal effects. However, gas
present in RIAFs is hot and fully ionized and ambipolar diffusion in high
ionization state is negligible. Furthermore, the mean free path for the
Coulomb interactions in hot accretion flows is larger than the typical
length-scale of the system, the Coulomb coupling between ions and electrons
is not strong enough and such flows are effectively collisionless
\citep{Quataert1998}. Therefore, the gyration timescale in these flows
is much shorter than the dynamical timescale and the Hall effect can be
safely ignored \citep[see][]{Pandey2008}. Hence, inclusion of resistivity
is essential to trigger the magnetic reconnection
\citep[e.g.,][]{Ding2010,Ripperda2019ApJS,Ripperda2020}.

Much effort has been devoted to the study of resistivity in hot
accretion flows, and interesting results have been obtained. For example,
\cite{Zeraatgari2018} showed that the ohmic resistivity in such flows modifies
the amount of magnetic field and the role of the magnetic field in transferring
angular momentum. \cite{Abbassi2012ApSS} also showed that the
rotational velocity of the resistive accretion flows in the absence of
outflow depends on the magnetic diffusivity. They found that the effect
of magnetic diffusivity on the rotational velocity depends on the field
components \citep[see also][]{Ghoreyshi2020PASA}. In the presence of outflows,
the results indicate that the rotational velocity of hot accretion flows
is always sub-Keplerian and the disk reaches a non-rotating limit when
the magnetic diffusivity increases \citep{Ghoreyshi2020PASA}. We know that
the magnetic diffusivity not only modifies the structure of accretion flows,
but also can affect wind lunching and the strength of the disk wind
\citep{Kuwabara2000,Fendt2002,Igumenshchev2003,Cemeljic2014,Qian2018,Vourellis2019}.
On the other hand, the magnetic dissipation heating in regions of a disk
where outflows may play an important role is comparable to the viscous
heating \citep{Zeraatgari2018}. This heating mechanism helps the electron
to have a higher temperature \citep{Bisnovatyi1997,Ding2010}. Hence, the
magnetic diffusivity can be an important mechanism in hot accretion flows.

As mentioned above, the Coulomb collision mean free path of charged particles
(both electrons and ions) in the accretion flow of extremely LLAGNs, such
as Sgr A* and M 87, is much larger than the typical size of the system
\citep[see][]{Bu2019Uni}. In such weakly-collisional flows, anisotropic
thermal conduction can be significant and along the magnetic field lines
\citep{Parrish2007,Sharma2008,Bu2011}. The anisotropic thermal conduction
plays an important role in the transport of thermal energy from the inner
(hotter) to the outer (cooler) parts. Thus, the anisotropic thermal conduction
affects the properties of outflow such as the energy flux
\citep{Bu2016MNRAS}. Since the ions' mean free path in weakly-collisional
accretion flows is larger than its Larmor radius, the pressure perpendicular
to the magnetic field line and that parallel to the magnetic field line
are not the same \citep{Quataert2002,Sharma2003,Chandra2015}. Previous
findings indicate that the properties of a hot accretion flow depend significantly
on anisotropic pressure. For example, the mass inflow rate in hot accretion
flows with the anisotropic pressure decreases towards the central object
when a very weak magnetic field is present \citep{Wu2017}. By considering
a relatively strong magnetic field, \cite{Bu2019Uni} found that the infall
velocity increases with the anisotropic pressure and with the outflows.
We know that the changes in the infall velocity due to the outflow are
dependent on the magnetic diffusivity when the anisotropic pressure is
absent \citep{Ghoreyshi2020PASA}. To the best of our knowledge, the effect
of magnetic diffusivity on the properties of hot accretion flows with the
anisotropic pressure has not yet been studied in detail. Here, we will
study the dynamics of resistive hot accretion flows in the presence of
anisotropic pressure. Indeed, our main aim is to explore the role of magnetic
diffusivity and compare the resistive heating with the other heating mechanisms.

Here, the magnetic field is assumed to have only a $\phi $-component. We
present the self-similar solutions of resistive hot accretion flows when
the outflows are present. The paper is organized as follows. We first formulate
basic equations for such a disk in Sect.~2. In Sect.~3, the basic equations
are solved using self-similar method and the results are discussed in Sect.~4. We then summarize our conclusions in the final section.

\section{Basic Equations}\label{BasicEquations}

In this paper, we study the time-steady ($\partial /\partial t=0$) and
the axisymmetric ($\partial /\partial \phi =0$) resistive hot accretion
flows in the cylindrical coordinates $(r, \phi , z)$. All flow variables
are assumed to depend only on $r$. The basic equations in Gaussian units
are as follows:
\begin{equation}\label{eq:continuity1}
\frac{d\rho}{dt}+\rho\nabla\cdot \mathbf{v}=0,~~~~~~~~~~~~~~~~~~~~~~~~~~~~~
\end{equation}

\begin{eqnarray}\label{eq:momentum}
\rho \frac{d \mathbf{V}}{dt}=-\nabla p-\rho\nabla\Phi+\frac{1}{4\pi}(\nabla\times\mathbf{B})\times\mathbf{B}~~~~~
\nonumber\\~~~+\nabla\cdot \mathbf{T}+\nabla\cdot \mathbf{\Pi},
\end{eqnarray}

\begin{equation}\label{eq:energy1}
\rho \big(\frac{d e}{dt}-\frac{p}{\rho^2}\frac{d\rho}{dt}\big)=(q_{\rm vis}+q_B)-q_{\rm rad},~~~~~~~~~~~~~~~~~~~~
\end{equation}

\begin{equation}\label{induction}
\frac{\partial\mathbf{B}}{\partial t}=\nabla\times(\mathbf{v}\times\mathbf{B}-\eta\nabla\times\mathbf{B}),~~~~~~~~~~~~~~~~~~~~
\end{equation}
where $d/dt=\partial /\partial t+\mathbf{v}\cdot \nabla $ is the Lagrangian
time derivative. Here, $\rho $ and $\mathbf{v}=(v_{r},v_{\phi },0)$ are the
midplane density of the disk and the velocity vector, respectively.
$p$ is the gas pressure and $\Phi $ is the gravitational potential. We
ignore relativistic effects and use Newtonian gravity ($\Phi =-GM_{*}/r$).
From the numerical MHD simulations done by \cite{Hirose2004}, the magnetic
field in the main disk body, corona, and inner torus is mainly toroidal.
But the poloidal component can be important in regions near the black hole.
Hence, we assume that the dominant field component in most regions of the
accretion flow is toroidal. In the present paper, therefore, a large-scale
magnetic field is adopted that has only a toroidal component
$B_{\phi }$. In Eq.~(\ref{induction}), $\eta $ is the magnetic diffusivity.

\begin{figure*}
\includegraphics{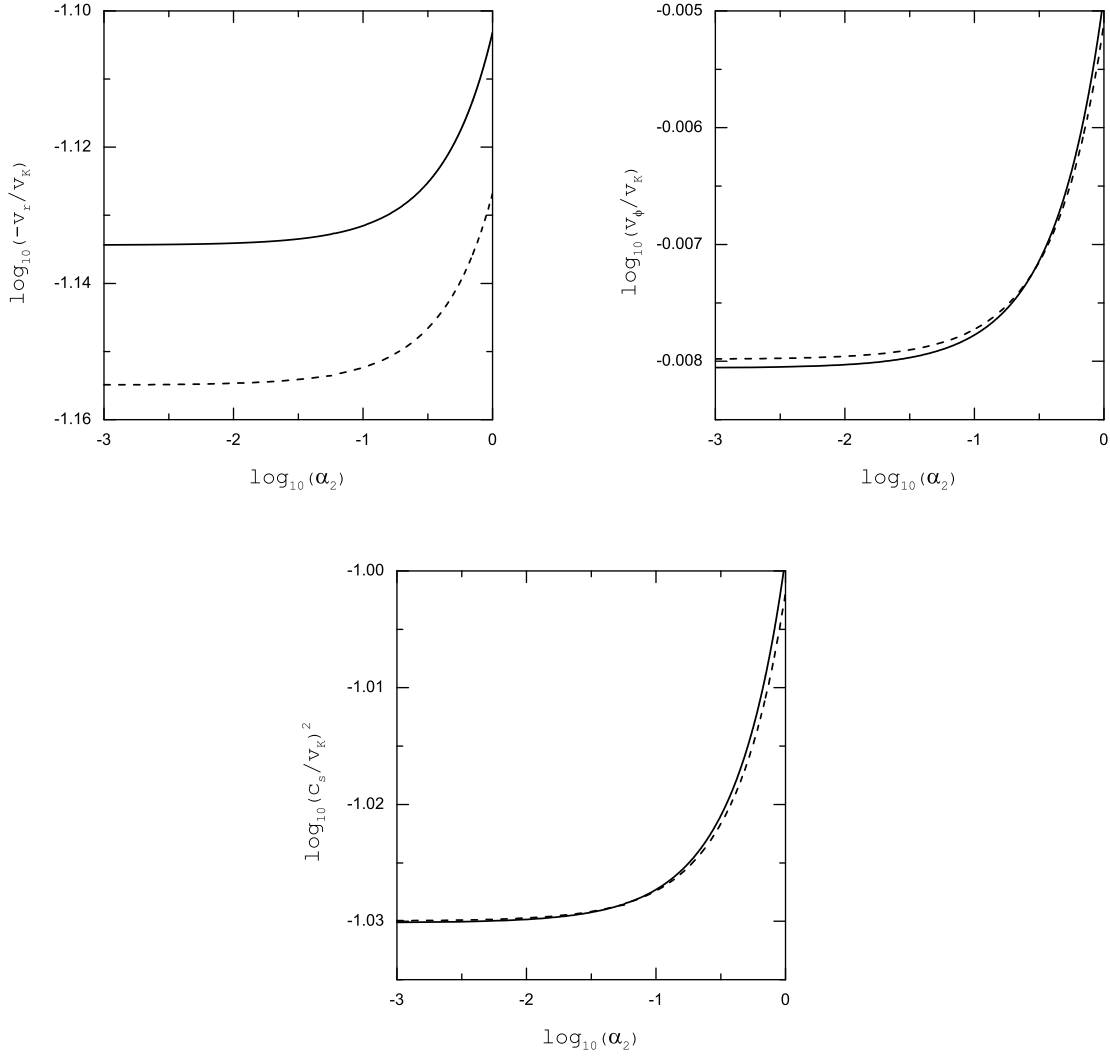}
\caption{Profiles of the physical variables of the accretion disk versus
the strength of anisotropic pressure. A comparison between two forms of
the viscosity $\nu $ for $\gamma =4/3$, $s=1/2$, $\alpha _{1}=0.1$,
$\beta _{\phi }=0.1$, $\zeta _{1}=\zeta _{2}=\zeta _{3}=1.2$. While the solid
curves correspond to the definition of $\nu =\alpha c_{s} H$, the dashed
curves show the results of $\nu =\alpha {c_{s}}^{2} /\Omega _{K} $}
\label{fig:f1}
\end{figure*}

The fourth term on the right side of Eq.~(\ref{eq:momentum}),
$\nabla \cdot \mathbf{T}$, represents the angular momentum transferred
by the turbulent magnetic field. If the $r\phi $-component of the viscous
stress tensor in an accretion disk is dominant, we have
\citep{Kato2008}
%
\begin{equation}
\label{Trphi}
T_{r\phi }=\rho \nu _{1} r (\frac{d\Omega }{dr}),~~~~~~~~~~~~~~~~~~~~
\end{equation}
where $\nu _{1}=\alpha _{1} c_{s} H$. Here, $H$ is disk half-thickness,
and $c_{s}=\sqrt{p/\rho }$ is sound speed. When the magnetic fields exist
in an accretion disk, both $c_{s}$ and $H$ are written as a function of
the magnetic field strength. Therefore, we cannot adopt the simple form
$\nu _{1}=\alpha c_{s}^{2}/\Omega _{K}$ for the viscosity. In our paper,
we will compare the results of these two forms of viscosity (see Fig.~\ref{fig:f1}).

The anisotropic pressure $\boldsymbol{\Pi}$ in Eq.~(\ref{eq:momentum}) which
can be denoted by an anisotropic viscosity is written as
\citep{Braginskii1965,Balbus2004,Chandra2015}
%
\begin{equation}
\label{Pi1}
\boldsymbol{\Pi}=-3\rho \nu _{2} \big [\mathbf{b}\mathbf{b}:\nabla
\mathbf{v}-\frac{\nabla \cdot \mathbf{v}}{3}\big ] \big [\mathbf{b}
\mathbf{b}-\frac{\mathbf{I}}{3}\big ].~~~~~~~~~~~~~~~~~~~~
\end{equation}
Here, $\mathbf{b}=\mathbf{B}/|\mathbf{B}|$ and $\mathbf{I}$ are the unit
vector in magnetic field direction and the unit tensor, respectively. The
anisotropic viscosity $\nu _{2}$ is also written as
$\alpha _{2} c_{s} H$ in which $\alpha _{2}$ represents the strength of
anisotropic pressure. When the magnetic field has only a toroidal component,
the non-zero components of the anisotropic pressure are as follows
\citep{Bu2019Uni},
%
\begin{equation}
\label{Pi2}
\Pi _{rr}=\Pi _{zz}=-\frac{1}{2}\Pi _{\phi \phi }=-
\frac{\rho \nu _{2} r^{2}}{3} \frac{d}{dr}\big (\frac{v_{r}}{r^{2}}
\big ),~~~~~~~~~~~~~~~~~~~~
\end{equation}
where $v_{r}$ is the radial infall velocity. Since the magnetic field geometry
modifies the anisotropic pressure tensor, the components of anisotropic
pressure can be very complicated in the presence of a poloidal magnetic
field. Therefore, we consider the dominant component of the magnetic field,
i.e., the toroidal component, for simplicity.

In Eq.~(\ref{eq:energy1}), $e$, $q_{\mathrm{vis}}$, $q_{B}$, and
$q_{\mathrm{rad}}$ represent the gas specific internal energy, the heating rate
of the gas, the resistive heating, and the radiative cooling rates, respectively.
We assume that the right side of Eq.~(\ref{eq:energy1}) is equal to
$f(q_{\mathrm{vis}}+q_{B})$ in which the advection factor can be in the range
$0\leq f \leq 1$. In this paper, we set $q_{\mathrm{rad}}=0$, and then the advection
factor $f=1.0$. The gas heating rate $q_{\mathrm{vis}}$ is the sum of the heating
by the magnetic reconnection due to the turbulent component of the magnetic
field ($q_{\mathrm{vis1}}$) and the heating by the work done by anisotropic
pressure ($q_{\mathrm{vis2}}$). These heating rates are defined by

\begin{center}
$q_{\mathrm{vis1}}=\rho \nu _{1} r^{2} \big (\frac{d\Omega }{dr}\big )^{2}$
\end{center}
and
\begin{center}
$q_{\mathrm{vis2}}=\frac{1}{3}\rho \nu _{2} r^{4} \big [\frac{d}{dr} (
\frac{v_{r}}{r^{2}})\big ]^{2}$,
\end{center}
where $\Omega =v_{\phi }/r$ is the angular velocity of the gas and
$v_{\phi }$ is the rotational velocity. The resistive heating rate is also
written as
\begin{center}
$q_{B}=\frac{\eta }{4\pi }\mid \nabla \times \mathbf{B}\mid ^{2}$.
\end{center}
The diffusivity is believed to be generated by turbulence triggered within
the accretion disk \citep{Zanni2007}. Hence, we assume that the anomalous
magnetic diffusivity $\eta $ can be expressed in the same way as in
\cite{Bisnovatyi1976}, namely,
\begin{center}
$\eta =\eta _{0} c_{s} H$,
\end{center}
where the $\alpha $-prescription of \cite{Shakura1973} is adopted and the
magnetic diffusivity parameter $\eta _{0}$ is a constant, less than unity.
Note that this form of the magnetic diffusivity was used in some previous
work \citep[e.g.,][]{Shadmehri2004,Abbassi2012ApSS,Samadi2014}, although $H$ was replaced by $c_{s}/\Omega _{K}$ in the definition of $\eta $.

According to the above relations, the radial and azimuthal momentum equations
and the energy equation in the presence of outflow are written as
%
\begin{eqnarray}
\label{eq:radial-momentum}
v_{r} \frac{d v_{r}}{dr}+\frac{1}{2\pi r\Sigma }\frac{d\dot{M}}{dr}
\big (v_{rw}-v_{r}\big )=\frac{{v_{\phi }}^{2}}{r}-\frac{GM_{*}}{r^{2}}~~~
\nonumber
\\
-\frac{1}{\Sigma }\frac{d}{dr}(\Sigma c_{s}^{2})-\frac{1}{2\Sigma }
\frac{d}{dr}\Big (\Sigma {c_{\phi }}^{2}\Big )-\frac{{c_{\phi }}^{2}}{r}~~~~~
\nonumber
\\
~~~~~+\frac{2v_{r}}{3r^{2}\Sigma }\Big [\nu _{2}\Sigma +\frac{d}{dr}
\big (r\nu _{2}\Sigma \big )\Big ]~~~~
\nonumber
\\
~~~-\frac{1}{3r\Sigma }\frac{d}{dr}\big (r\nu _{2}\Sigma
\frac{dv_{r}}{dr}\big ),
\end{eqnarray}
%
\begin{eqnarray}
\label{eq:azimuthal-momentum}
\frac{\Sigma v_{r}}{r} \frac{d}{dr} (r v_{\phi })+
\frac{1}{2\pi r\Sigma }\frac{d\dot{M}}{dr}\big (v_{\phi w}~~~~~~~~~~~~~~
\nonumber
\\
~~~~~~~~~~~~~~~~~~-v_{\phi }\big )=\frac{1}{r^{2}}\frac{d}{dr}( r^{3}
\nu _{1}\Sigma \frac{d\Omega }{dr}),
\end{eqnarray}
%
\begin{eqnarray}
\label{eq:energy}
\frac{v_{r}}{\gamma -1}\frac{d c_{s}^{2}}{dr}-
\frac{v_{r} c_{s}^{2}}{\rho }\frac{d\rho }{dr}+\frac{1}{2\pi r\Sigma }
\frac{d\dot{M}}{dr}\big (e_{w}-e\big )
\nonumber
\\
~~~=f\Bigg \{ r^{2}\nu _{1}\big (\frac{d\Omega }{dr}\big )^{2}+
\frac{1}{3}r^{4}\nu _{2}\bigg [\frac{d}{dr}(\frac{v_{r}}{r^{2}})\bigg ]^{2}
\nonumber
\\
~~~~~~~~~~~~+\frac{\eta }{4\pi \rho }\bigg [\frac{1}{r}\frac{d}{dr}(r B_{\phi })\bigg ]^{2} \Bigg \},
\end{eqnarray}
where the surface density $\Sigma $ and the Alfv\'{e}n sound speed
$c_{\phi }$ are defined as $\Sigma =2\rho H$, and
${c_{\phi }}^{2}={B_{\phi }}^{2}/ 4\pi \rho $. $v_{rw}$ and $v_{\phi w}$ denote
the $r$- and $\phi $-components of the velocity of the outflow.
$e_{w}$ is the specific internal energy of the outflow. In the present
paper, we assume $v_{rw}=\zeta _{1} v_{r}$,
$v_{\phi w}=\zeta _{2} v_{\phi }$, and $e_{w}=\zeta _{3} e$
\citep{Bu2009}.

The hydrostatic balance in the vertical direction leads to a relation between
$H$ and $c_{s,\phi }$. We have
%
\begin{equation}
\label{eq:hydrostatic}
\frac{GM_{*}}{r^{3}}H^{2}={c_{s}}^{2}(1+\beta _{\phi }),~~~~~~~~~~~~~~~~~~~~~
\end{equation}
where $\beta _{\phi }$ is the ratio of the magnetic pressure to the gas pressure,
i.e., $\beta _{\phi }=(1/2)(c_{\phi }/c_{s})^{2}$, which can describe the strength
of the large-scale magnetic field. For hot accretion flows,
$\beta _{\phi }$ may lie in the range from 0.01 to 1
\citep[e.g.][]{DeVilliers2003,Beckwith2008}.


\section{Self-similar solutions}

As mentioned earlier, the variables of the disk are assumed to be as a
function of $r$. We use self-similar solutions in the following forms
\citep{Narayan1994}:
%
\begin{equation}
\label{eq:v}
v_{r}(r)=-C_{1} \alpha _{1} v_{K},~~~~~~~~~~~~~~~~
\end{equation}
%
\begin{equation}
\label{eq:o}
v_{\phi }(r)=C_{2} v_{K},~~~~~~~~~~~~~~~~~
\end{equation}
%
\begin{equation}
\label{eq:cs}
c_{s}^{2}(r)=C_{3}{v_{K}}^{2}.~~~~~~~~~~~~~~~~~~~~~~~~~~~~~
\end{equation}
By using the form $\Sigma =\Sigma _{0} r^{s}$ for the surface density and
the definition of $\dot{M}$ ($=-2\pi r \Sigma v_{r}$), the accretion rate
takes the form
\begin{center}
$\dot{M}=2\pi \alpha _{1} C_{1} \Sigma _{0}\sqrt{GM_{*}}r^{s+
\frac{1}{2}}$.
\end{center}
Here, $\Sigma _{0}$ and $s$ are two physical constants and $s$ is between
$-1/2$ and 1/2. We note that $s=-1/2$ corresponds to the cases without
outflow. By substituting these self-similar solutions into the basic Eqs.~(\ref{eq:radial-momentum})--(\ref{eq:energy}) and using Eq.~(\ref{eq:hydrostatic}),
we have
%
\begin{eqnarray}
\label{eq:radial-momentum1}
-\Big [\frac{1}{2}+(s+\frac{1}{2})(\zeta _{1}-1)\Big ]{\alpha _{1}}^{2} {C_{1}}^{2}={C_{2}}^{2}-1
\nonumber
\\
-2\beta _{\phi }C_{3}-(s-1)(1+\beta _{\phi })C_{3}
\nonumber
\\
~~~~~~-\frac{5}{6}\alpha _{1}\alpha _{2}\sqrt{1+\beta _{\phi }}(s+2) C_{1}
C_{3},
\end{eqnarray}
%
\begin{equation}
\label{eq:azimuthal-momentum1}
\Big [1-(2s+1)(\zeta _{2}-1)\Big ]C_{1}=3\sqrt{1+\beta _{\phi }}(s+1)C_{3},
\end{equation}
%
\begin{eqnarray}
\label{eq:energy2}
\Big [\frac{1}{\gamma -1}+s-1+
\frac{(2s+1)(\zeta _{3}-1)}{2(\gamma -1)}\Big ]C_{1}=f~~~~~~~~~~
\nonumber
\\
~~~~\times \sqrt{1+\beta _{\phi }}\Big (\frac{9}{4}{C_{2}}^{2}+
\frac{25}{12}\alpha _{1}\alpha _{2}{C_{1}}^{2} +
\frac{\beta _{\phi }\eta _{0}s^{2}}{2\alpha _{1}}C_{3}\Big ).
\end{eqnarray}
Adopting an odd function of $z$ for $B_{\phi }$ and a simple form of the
viscosity, Eqs.~(\ref{eq:radial-momentum1})--(\ref{eq:energy2}) in the
absence of magnetic diffusivity reduce to Eqs.~(19), (20), and (22) of
\cite{Bu2019Uni}. If the parameters $s$, $\alpha _{1}$,
$\alpha _{2}$, $\beta _{\phi }$, $\eta _{0}$, $\zeta _{1}$,
$\zeta _{2}$, $\zeta _{3}$, and $\gamma $ have given values, there is a
closed set of equations for $C_{1}$, $C_{2}$, and $C_{3}$ to determine
the variables of the accretion flow. By using the above equations, a second-order
algebraic equation for $C_{1}$ is obtained as follows:
%
\begin{eqnarray}
\label{eq:main-equation}
\Bigg \{
\frac{5(s+2)\bigg [1+(2s+1)(1-\zeta _{2})\bigg ]\alpha _{1}\alpha _{2}}{8(s+1)}~~~~~~~~~~~~~~
\nonumber
\\
~~~-\frac{9\bigg [1+(2s+1)(\zeta _{1}-1)\bigg ]{\alpha _{1}}^{2}}{8}+
\frac{25\alpha _{1}\alpha _{2}}{12}\Bigg \}{C_{1}}^{2}
\nonumber
\\
+\Bigg \{-\frac{1}{f\sqrt{1+\beta _{\phi }}}\bigg [s-1+
\frac{(2s+1)(\zeta _{3}-1)}{2(\gamma -1)}~~
\nonumber
\\
~~~~~~~+\frac{1}{\gamma -1}\bigg ]+\bigg [
\frac{1+(2s+1)(1-\zeta _{2})}{2(s+1)}\bigg ]
\nonumber
\\
\times \bigg [\frac{3\beta _{\phi }}{\sqrt{1+\beta _{\phi }}}+
\frac{3(s-1)\sqrt{1+\beta _{\phi }}}{2}
\nonumber
\\
~~~~
\frac{\beta _{\phi }\eta _{0} s^{2}}{3\sqrt{1+\beta _{\phi }}\alpha _{1}}
\bigg ]\Bigg \}C_{1}+\frac{9}{4}=0.
\end{eqnarray}

Noting that the resistive heating rate is proportional to
$\beta _{\phi }\eta _{0} s^{2}$ (see the last term on the right side of Eq.~(\ref{eq:energy2})); we will discuss the role of each of these parameters.
Here, we focus on the effects $\eta _{0}$, $\beta _{\phi }$, $s$, and
$\alpha _{2}$. The other parameters are assumed to be constant;
$\gamma =4/3$, $f=1$, $\alpha _{1} =0.1$, and
$\zeta _{1}=\zeta _{2}=\zeta _{3}=1.2$. By setting these parameters, Eqs.~(\ref{eq:radial-momentum1})--(\ref{eq:energy2}) lead to the dependence
of $C_{2}$ and $C_{3}$ on $C_{1}$ as follows:
%
\begin{eqnarray}
\label{eq:c2}
C_{2}=\Big [1-
\frac{\sqrt{1+\beta _{\phi }}(0.8-0.4s)(1-s)C_{1}}{3(s+1)}~~~~~~~~~~~~
\nonumber
\\
~+ \frac{2\beta _{\phi }(0.8-0.4s)C_{1}}{3\sqrt{1+\beta _{\phi }}(s+1)}-0.005(1.2+0.4s){C_{1}}^{2}
\nonumber
\\
~+\frac{0.028\alpha _{2}(0.8-0.4s)(s+2){C_{1}}^{2}}{s+1}\Big ]^{1/2},
\end{eqnarray}
%
\begin{equation}
\label{eq:c3}
C_{3}=\frac{(0.8-0.4s)}{3\sqrt{1+\beta _{\phi }}(s+1)}C_{1}.
\end{equation}
Here, $C_{1}$ is obtained by solving Eq.~(\ref{eq:main-equation}).
\section{Results}

In order to examine the effects of anisotropic pressure on the properties
of a resistive hot accretion flow, we obtain the variables $C_{1}$,
$C_{2}$, and $C_{3}$ as a function of the strength of anisotropic pressure
$\alpha _{2}$. In this paper, the strength of anisotropic pressure is assumed
to be in the range from 0.001 to 1.0. The other input parameters are
$\eta _{0}=0.1$, $s=0.5$, and $\beta _{\phi }=0.1$, unless stated otherwise.
Since \cite{Bu2019Uni} studied the effect of parameters $\zeta _{1}$,
$\zeta _{2}$, and $\zeta _{3}$ on the disk properties, these parameters,
as mentioned above, are assumed to be fixed. We also note that only real
roots corresponding to positive ${C_{2}}^{2}$ must be adopted.

\begin{figure*}
\includegraphics{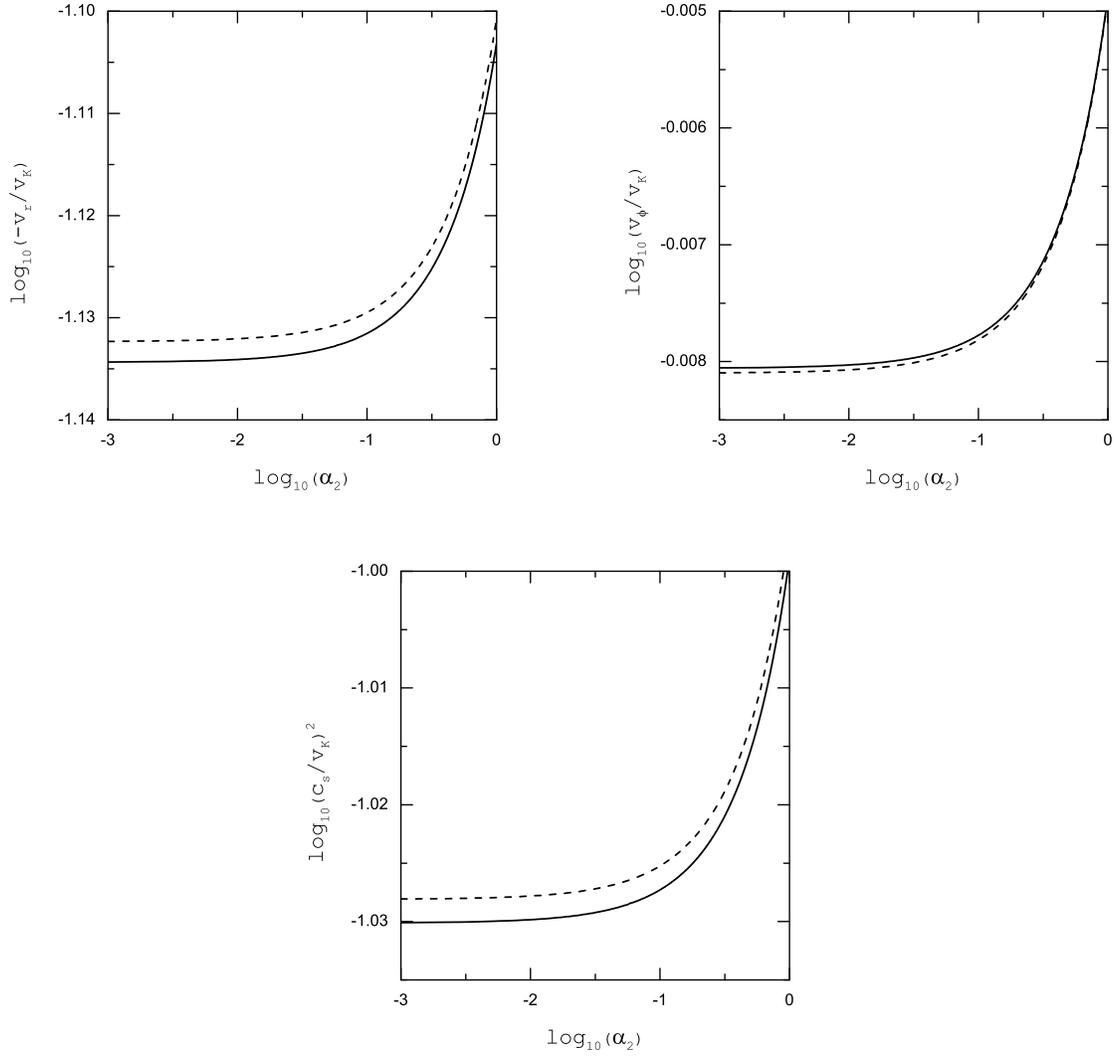}
\caption{Profiles of the physical variables of the accretion disk versus
the strength of anisotropic pressure for different values of
$\eta _{0}$. The solid and the dashed curves correspond to
$\eta _{0}= 0.1$ and 1.0, respectively}
\label{fig:f2}
\end{figure*}

As mentioned previously, the viscosity in the magnetized disk does not
obey the form $\nu =\alpha {c_{s}}^{2}/\Omega _{K}$. According to
$\nu =\alpha c_{s} H$ and Eq.~(\ref{eq:hydrostatic}), the viscosity in
the magnetized disk is defined as
$\nu =\alpha \sqrt{1+\beta _{\phi }} {c_{s}}^{2}/\Omega _{K}$. For now, we
will compare the viscosities $\nu =\alpha c_{s} H$ and
$\nu =\alpha {c_{s}}^{2}/\Omega _{K}$. In Fig.~\ref{fig:f1}, the results
of this comparison are illustrated. Here, we assume that $s=1/2$,
$\alpha _{1}=0.1$, $\beta _{\phi }=0.1$,
$\zeta _{1}=\zeta _{2}=\zeta _{3}=1.2$. All profiles are displayed as a
function of the strength of anisotropic pressure $\alpha _{2}$. We find
that the disk variables effectively are enhanced as the strength of anisotropic
pressure increases. The strength of the anisotropic pressure is one of
the important parameters in the heating by anisotropic pressure. Since
the sound speed is proportional to $T^{1/2}$, one can expect that the sound
speed increases with the strength of the anisotropic pressure. The last
term on the right-hand side of Eq.~(\ref{eq:radial-momentum1}) is the force
of the divergence of the anisotropic pressure. For our input parameters,
this force is always negative. This means that the direction of the force
of the divergence of the anisotropic pressure is towards the central object.
According to Eq.~(\ref{eq:radial-momentum1}), an increase in the strength
of the anisotropic pressure causes this force, and thus the radial infall
velocity, to increase \citep[see also][]{Bu2019Uni}. By solving Eqs.~(\ref{eq:radial-momentum1})--(\ref{eq:energy2}),
we find that the rotational velocity is an increasing function of
$\alpha _{2}$ (see Eq.~(\ref{eq:c2})). This increasing trend is seen in
the top right panel of Fig.~\ref{fig:f1}.

\begin{figure*}
\includegraphics{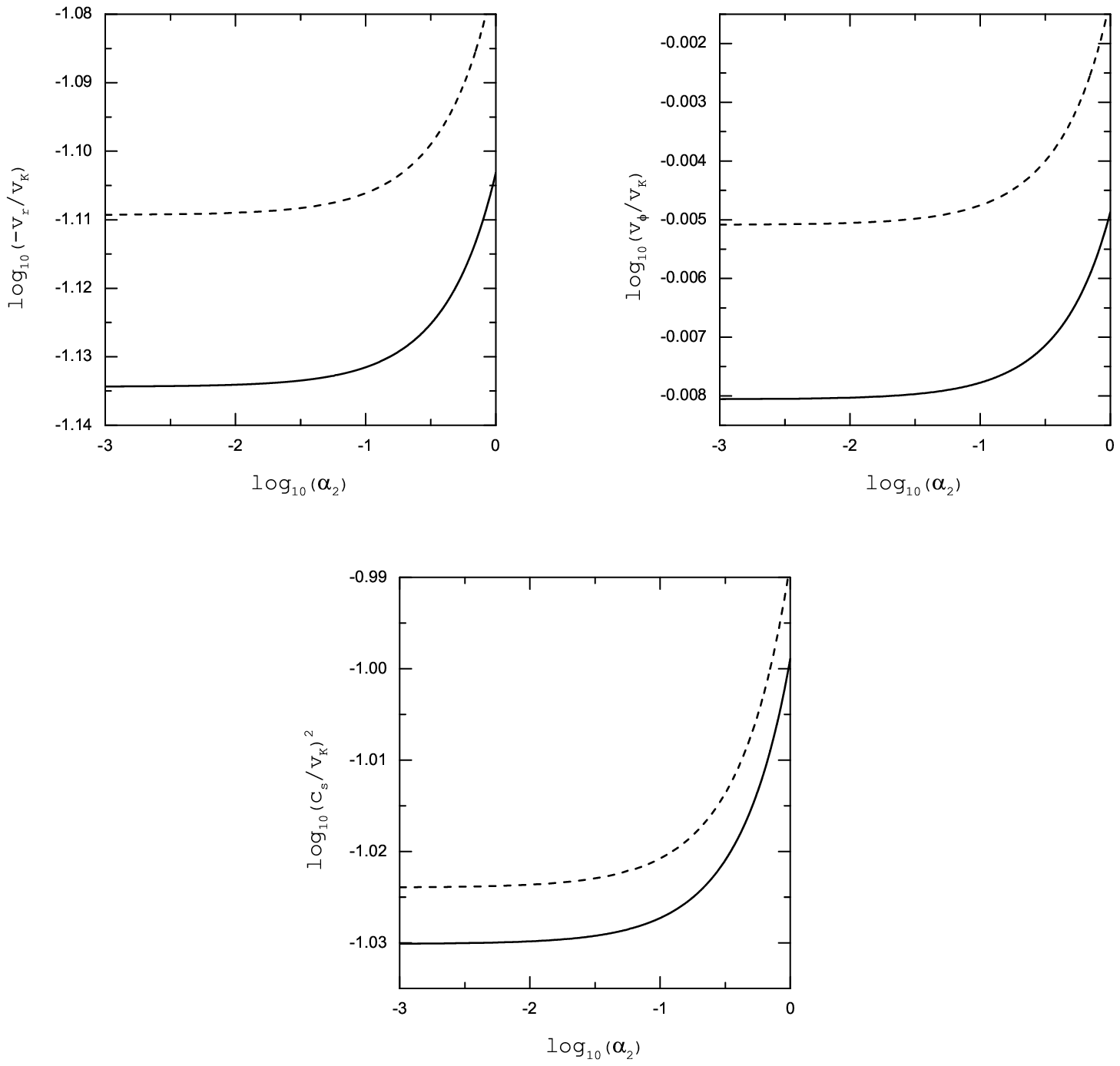}
\caption{Similar to Fig.~\protect\ref{fig:f2}. Profiles of the physical
variables of the accretion disk versus the strength of anisotropic pressure
for different values of $\beta _{\phi }$ when $\eta _{0}= 0.1$. The solid
and the dashed curves correspond to $\beta _{\phi }= 0.1$ and 0.2, respectively}
\label{fig:f3}
\end{figure*}

\begin{figure*}
\includegraphics{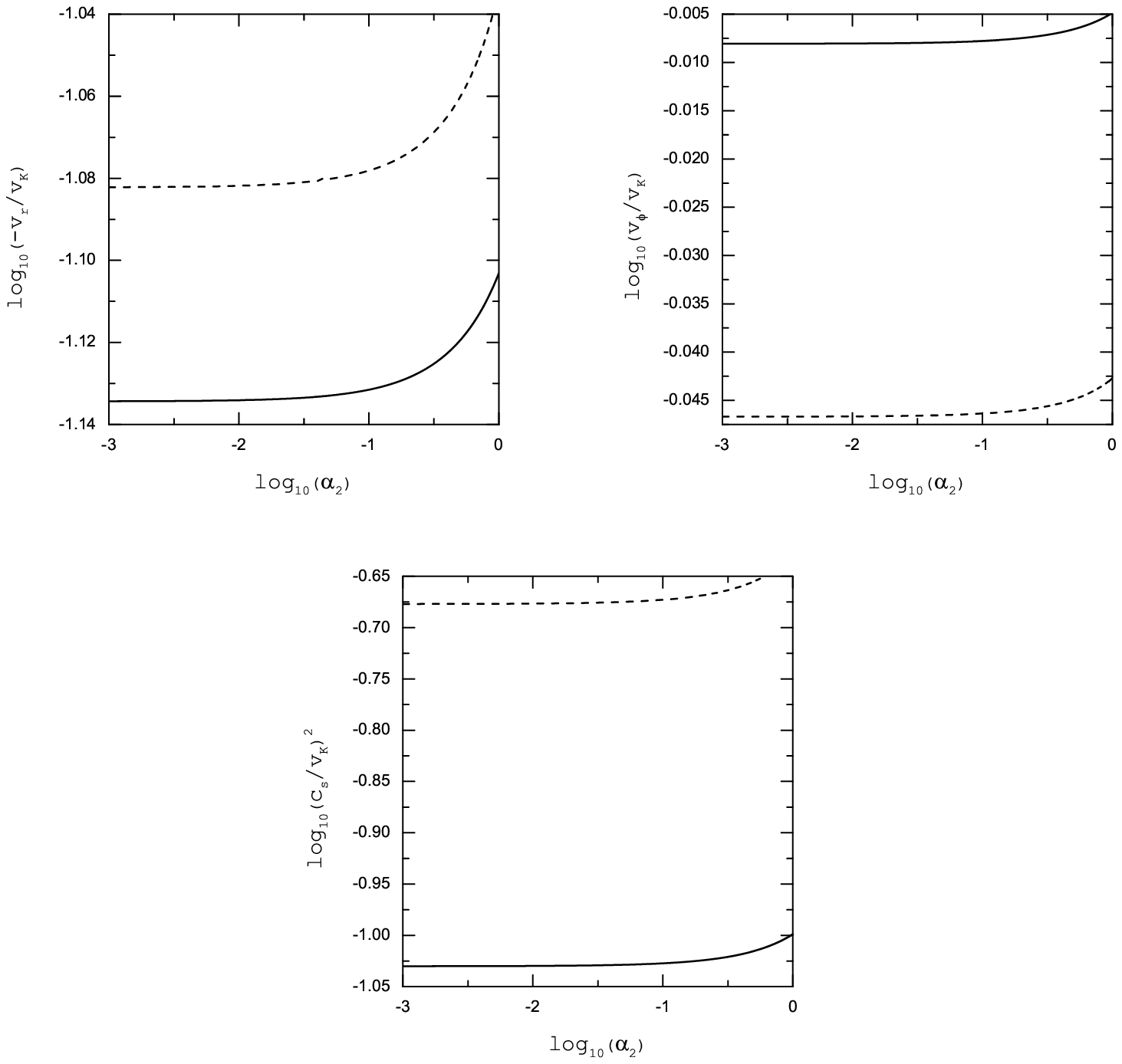}
\caption{Similar to Fig.~\protect\ref{fig:f2}. Profiles of the physical
variables of the accretion disk versus the strength of anisotropic pressure
for different values of $s$ when $\eta _{0}= 0.1$. The solid and the dashed
curves correspond to $s= 0.5$ and 0.0, respectively}
\label{fig:f4}
\end{figure*}

\begin{figure*}
\includegraphics{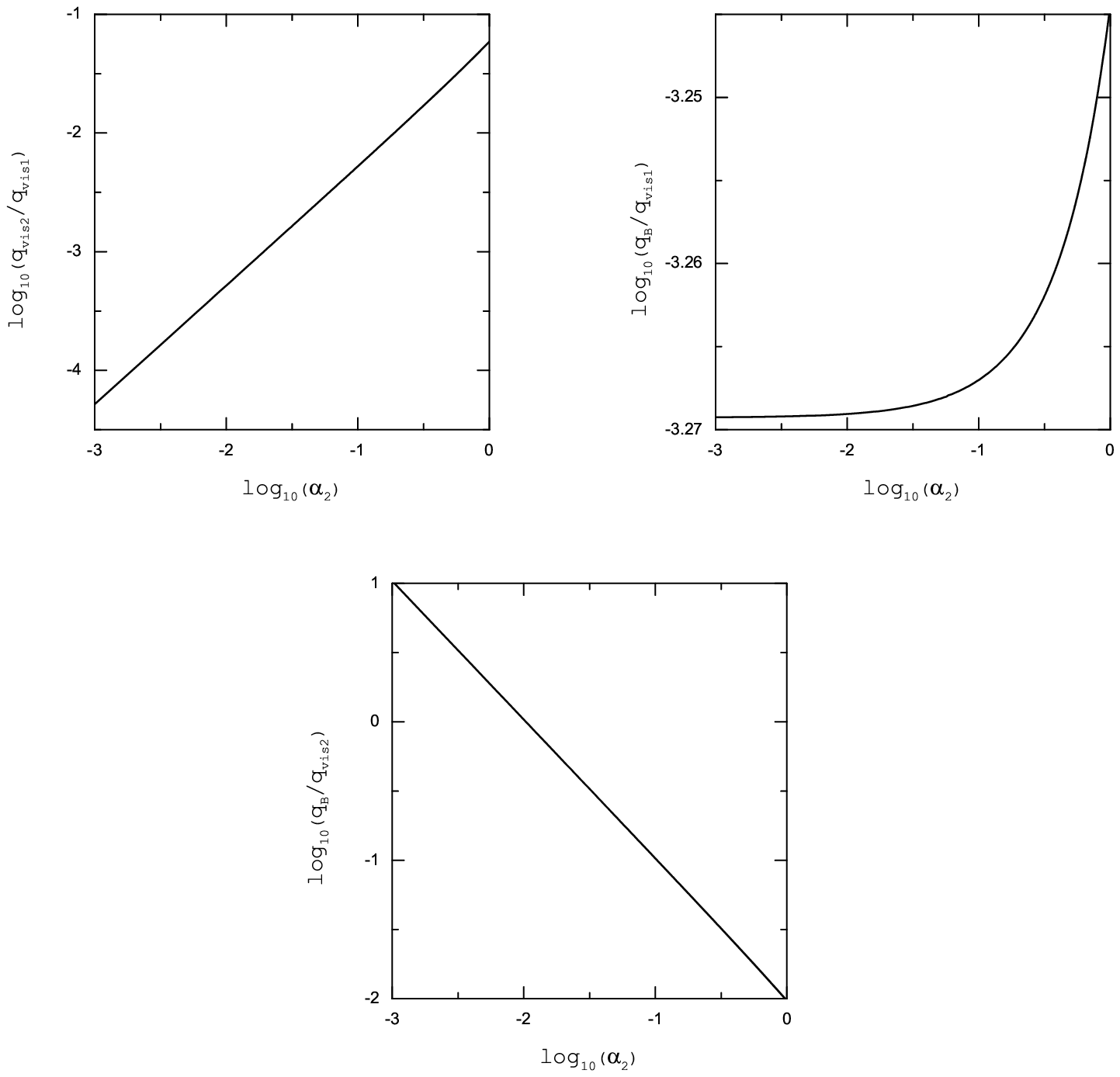}
\caption{Profiles of the ratio of heating rates as a function of the strength
of anisotropic pressure. We adopt $\eta _{0}=0.1$,
$\beta _{\phi }=0.1$, and $s=0.5$}
\label{fig:f5}
\end{figure*}

The top left panel in Fig.~\ref{fig:f1} is the normalized radial infall
velocity ($v_{r}/v_{K}$). We find that the modified viscosity
$\nu =\alpha \sqrt{1+\beta _{\phi }} {c_{s}}^{2}/\Omega _{K}$ increases the
infall velocity. One can expect that the value of $\beta _{\phi }$ plays
a key role in these changes \citep[see also][]{Zhang2008}. Although this
increase occurs within all $\alpha _{2}$, the strength of the anisotropic
pressure affects these changes. When $\alpha _{2}$ rises, the increase
in infall velocity is more significant. The modification of the viscosity
does not significantly change the rotational velocity (top right panel)
and the sound speed (bottom panel). In the high-$\alpha _{2}$ limit, these
velocities increase due to the modified viscosity. By reducing
$\alpha _{2}$, the modified viscosity causes the rotational velocity and
the sound speed to reduce. Our results in the low-$\alpha _{2}$ limit are
in good agreement with the findings of \cite{Zhang2008}. They also pointed
out that the modified viscosity decreases the radial velocity and the sound
speed. We will apply the modified viscosity form
$\nu =\alpha \sqrt{1+\beta _{\phi }} {c_{s}}^{2}/\Omega _{K}$ to the profiles
in the next figures.

The influence of magnetic diffusivity parameter $\eta _{0}$ is displayed
in Fig.~\ref{fig:f2}. Here, the adopted values of the magnetic diffusivity
parameter are 0.1 (solid curves) and 1.0 (dashed curves). Noting that the
resistive heating is directly related to the $\eta _{0}$, an increase in
the magnetic diffusivity parameter causes the disk to have a higher temperature.
Therefore, the sound speed increases with the magnetic diffusivity parameter.
According to Eq.~(\ref{eq:c3}), the radial velocity also rises. In the
absence of anisotropic pressure, \cite{Ghoreyshi2020PASA} also suggested
that the radial velocity and the sound speed increase with the magnetic
diffusivity. In the present paper, we find that these results depend on
the strength of anisotropic pressure. In the low-$\alpha _{2}$ limit, the
influence of magnetic diffusivity parameter is more significant. We also
find that an increase in the magnetic diffusivity parameter causes the
disk to rotate with a slower rate (see Eq.~(\ref{eq:c2})).

In Fig.~\ref{fig:f3}, we display the role of the ratio of the magnetic
to gas pressures $\beta _{\phi }$ for flows with $\eta _{0}=0.1$. The solid
and the dashed curves are the cases with $\beta _{\phi }= 0.1$ and 0.2, respectively.
Other input parameters are similar to Fig.~\ref{fig:f2}. According to Eq.~(\ref{eq:energy2}), the assumed heating rates depend directly on
$\beta _{\phi }$. The dependence of the heating rates on the ratio of the
magnetic to gas pressures causes the disk temperature to increase with
$\beta _{\phi }$, and thus the sound speed to be enhanced. One can see that
the anisotropic stress depends on the ratio of the magnetic to gas pressures
$\beta _{\phi }$ (see Eq.~(\ref{eq:radial-momentum1})). Particle-in-cell
simulations of the magnetorotational instability (MRI) in collisionless
plasmas have also indicated that the importance of the anisotropic stress depends on this ratio
\citep{Riquelme2012}. The signs of the magnetic force and of the force
of the divergence of the anisotropic pressure in Eq.~(\ref{eq:radial-momentum1})
are negative (i.e., they are directed towards the central black hole).
Thus, an increase in $\beta _{\phi }$ causes the infall velocity to rise
(see also Eq.~(\ref{eq:c3})). The dependence of the rotational velocity
on $\beta _{\phi }$ and $C_{1}$ implies that the disks with stronger magnetic
field rotate in a faster rate (see Eq.~(\ref{eq:c2})). These results are
in excellent agreement with previous work
\citep[e.g.,][]{Zhang2008,Ghoreyshi2020,Ghoreyshi2020PASA}.\looseness=1

Now, we explore the effects of $s$ on the physical variables of hot accretion
flows when $\eta _{0}=0.1$. In Fig.~\ref{fig:f4}, we set $s=0.5$ (solid
curves) and $s=0.0$ (dashed curves). The other parameters are similar to
Fig.~\ref{fig:f2}. In this paper, we consider the case for which the specific
velocities and internal energy of outflow are different from those of inflow
($\zeta _{i}s\neq 1$). Thus, the internal energy can be taken away from
the disk by the outflow. When outflows are stronger, i.e., higher
$s$, the value of $d\dot{M}/dr$ becomes higher. Hence, higher values of
the internal energy are extracted and the disk temperature falls strongly.
The reduction of the disk temperature leads to a decrease in the sound
speed and in the infall velocity. From Eq.~(\ref{eq:c2}), we see that a
decrease in the infall velocity enhances the rotational velocity.

As the next step toward a more complete description of resistive disks,
we display the ratio of different heating rates as a function of
$\alpha _{2}$. The top left panel of Fig.~\ref{fig:f5} shows the ratio
$q_{\mathrm{vis2}}/q_{\mathrm{vis1}}$. We find that the increase in
$\log (q_{\mathrm{vis1}}/q_{\mathrm{vis1}})$ with $\log (\alpha _{2})$ is linear
(with slope 1.03). One can see that the heating due to the anisotropic
pressure $q_{\mathrm{vis2}}$ in the low-$\alpha _{2}$ limit is negligible compared
to $q_{\mathrm{vis1}}$. However, the heating rate $q_{\mathrm{vis2}}$ can be significant
in the opposite limit. The ratio $q_{\mathrm{B}}/q_{\mathrm{vis1}}$ is illustrated
in the top right panel of Fig.~\ref{fig:f5}. A good fit to the profile
shown in this panel over a wide range of $\alpha _{2}$-values is an exponential
function. In the top panels of Fig.~\ref{fig:f5}, one can see that the
heating rates $q_{\mathrm{B}}$ and $q_{\mathrm{vis2}}$ increase with the strength
of the anisotropic pressure. But these two panels show that
$q_{\mathrm{vis1}}$ is a dominant heating rate over the two heating rates. On
the other hand, our results indicate that the profile of
$\log (q_{\mathrm{B}}/q_{\mathrm{vis2}})$ as a function of
$\log (\alpha _{2})$ is linear. Although the profile of
$\log (q_{\mathrm{vis1}}/q_{\mathrm{vis1}})$ has a positive slope, the slope of
$\log (q_{\mathrm{B}}/q_{\mathrm{vis2}})$ is negative with a value of about $-$1.02.
In low-$\alpha _{2}$ limit, the resistive heating is more significant than
the heating due to the anisotropic pressure (a factor of 10). Thereby,
the role of magnetic diffusivity is more important in low-$\alpha _{2}$
limit (see also Fig.~\ref{fig:f2}).

\section{Summary and discussion}

In hot accretion flows, the ions' mean-free path is larger than its Larmor
radius. Under these conditions, the pressure parallel to the magnetic field
and that perpendicular to the magnetic field are not the same. In such
collisionless flows, the pressure anisotropies may be produced by the amplification
of the magnetic field by MRI \citep{Sharma2006,Riquelme2012}. Simulations
done by \cite{Sharma2006} have indicated that these anisotropies can be
an important agent for angular momentum transport and particle heating
\citep[see also][]{Sharma2007}. On the other hand, the magnetic diffusivity
also plays a key role in such flows. In this paper, therefore, we studied
the dynamics of resistive hot accretion flows with the anisotropic pressure.
The toroidal field component was adopted and a modified viscosity form
was introduced in Sect.~\ref{BasicEquations}. We considered a steady state
flow and presented the self-similar solutions described by a function of
the radial distance.

We found that the modified viscosity form leads to infall velocities which
are quite different from the results of \cite{Bu2019Uni}. Our self-similar
solutions in the presence of outflows indicated that the magnetic diffusivity
causes the infall velocity to increase. Other parameters, such as the strength
of the anisotropic pressure and the ratio of the magnetic to gas pressures,
can also play a key role in the modification of the infall velocity. In
the high-$\alpha _{2}$ limit, for example, the radial velocity of inflow
rises strongly. Simulations of MRI have also demonstrated that the anisotropic
pressure in collisionless plasmas causes the angular momentum transport,
and thus the radial velocity, to be enhanced. On the other hand,
\cite{Kempski2019} pointed out that the angular momentum transport depends
not only on the anisotropic pressure, but also on the magnetic dissipation.
Since the disk material in the presence of these mechanisms is able to
move inward with faster rate, one can expect that the accretion rate of
a hot accretion flow is also modified. The change in accretion rate can
alter the black hole luminosity \citep{Bu2019ApJ}. A significant caveat
here is that we did not consider the poloidal magnetic field. Previous
work showed that in the ideal MHD the vertical component of the magnetic
field reduces the velocities discussed in this paper
\citep[e.g.,][]{Mosallanezhad2013}. However, the effect of the vertical
field component on the resistive disks structure depends on the magnetic
diffusivity \citep{Ghoreyshi2020PASA}. Then accounting for this component
may yield different results.

We found that the rotational velocity and the sound speed are almost unchanged
by the modified viscosity form. The rotational velocity changes insensitively
with the magnetic diffusivity parameter. Our results showed that the input
parameters such as $\alpha _{2}$, $\beta _{\phi }$, and $s$ can modify the
rotational velocity. When outflows are stronger, the sound speed decreases.
\cite{Bu2019Uni} also found that stronger outflows lead to a decrease in
the sound speed. In the presence of magnetic diffusivity and anisotropic
pressure, however, the disk temperature increases. An increase in the disk
temperature can lead to heating and acceleration of the electrons. Simulations
in collisionless plasmas have also showed that the viscous heating associated
with the anisotropic pressure can be an additional mechanism for the heating
of electrons \citep{Sharma2007,Riquelme2012,Kempski2019}. The heated and
accelerated electrons may be the origin of observed fluctuations (flares)
in the infrared and X-ray bands of Sgr A*. \cite{Ding2010} also proposed
that the resistive heating may be helpful in heating of the electrons.
We also suggest that the magnetic diffusivity and the anisotropic pressure
can help to explain non-thermal emission in the form of flares in Sgr A*.

\section*{Acknowledgements}
The authors would like to thank an unknown referee for a constructive report. The authors thank Dr. Mark Wardle for a helpful comment.

\nocite{*}
\bibliographystyle{spr-mp-nameyear-cnd}
\bibliography{myreference}

\begin{thebibliography}{81}
\ifx \bisbn   \undefined \def \bisbn  #1{ISBN #1}\fi
\ifx \binits  \undefined \def \binits#1{#1} \fi
\ifx \bauthor  \undefined \def \bauthor#1{#1} \fi
\ifx \batitle  \undefined \def \batitle#1{#1} \fi
\ifx \bjtitle  \undefined \def \bjtitle#1{#1}\fi
\ifx \bvolume  \undefined \def \bvolume#1{\textbf{#1}}\fi
\ifx \byear  \undefined \def \byear#1{#1} \fi
\ifx \bissue  \undefined \def \bissue#1{#1} \fi
\ifx \bfpage  \undefined \def \bfpage#1{#1} \fi
\ifx \blpage  \undefined \def \blpage #1{#1} \fi
\ifx \burl  \undefined \def \burl#1{\textsf{#1}} \fi
\ifx \doiurl  \undefined \def \doiurl#1{\textsf{#1}} \fi
\ifx \betal  \undefined \def \betal{\textit{et al.}} \fi
\ifx \binstitute  \undefined \def \binstitute#1{#1} \fi
\ifx \binstitutionaled  \undefined \def \binstitutionaled#1{#1} \fi
\ifx \bctitle  \undefined \def \bctitle#1{#1} \fi
\ifx \beditor  \undefined \def \beditor#1{#1} \fi
\ifx \bpublisher  \undefined \def \bpublisher#1{#1} \fi
\ifx \bbtitle  \undefined \def \bbtitle#1{#1} \fi
\ifx \bedition  \undefined \def \bedition#1{#1} \fi
\ifx \bseriesno  \undefined \def \bseriesno#1{#1} \fi
\ifx \blocation  \undefined \def \blocation#1{#1} \fi
\ifx \bsertitle  \undefined \def \bsertitle#1{#1} \fi
\ifx \bsnm \undefined \def \bsnm#1{#1} \fi
\ifx \bsuffix \undefined \def \bsuffix#1{#1} \fi
\ifx \bparticle \undefined \def \bparticle#1{#1} \fi
\ifx \barticle \undefined \def \barticle#1{#1} \fi
\ifx \bconfdate \undefined \def \bconfdate #1{#1} \fi
\ifx \botherref \undefined \def \botherref #1{#1} \fi
\ifx \url \undefined \def \url#1{\textsf{#1}} \fi
\ifx \bchapter \undefined \def \bchapter#1{#1} \fi
\ifx \bbook \undefined \def \bbook#1{#1} \fi
\ifx \bcomment \undefined \def \bcomment#1{#1} \fi
\ifx \oauthor \undefined \def \oauthor#1{#1} \fi
\ifx \citeauthoryear \undefined \def \citeauthoryear#1{#1} \fi
\ifx \endbibitem  \undefined \def \endbibitem {}\fi
\ifx \bconflocation  \undefined \def \bconflocation#1{#1} \fi
\ifx \arxivurl  \undefined \def \arxivurl#1{\textsf{#1}} \fi

\bibitem[\protect\citeauthoryear{{Abbassi} and
  {Mosallanezhad}}{2012}]{Abbassi2012ApSS}
\begin{barticle}
\bauthor{\bsnm{{Abbassi}}, \binits{S.}},
\bauthor{\bsnm{{Mosallanezhad}}, \binits{A.}}:
\bjtitle{\apss}
\bvolume{341}(\bissue{2}),
\bfpage{375}
(\byear{2012}).
\arxivurl{1205.3888}.
doi:\doiurl{10.1007/s10509-012-1147-x}
\end{barticle}
\endbibitem

\bibitem[\protect\citeauthoryear{{Balbus}}{2004}]{Balbus2004}
\begin{barticle}
\bauthor{\bsnm{{Balbus}}, \binits{S.A.}}:
\bjtitle{\apj}
\bvolume{616}(\bissue{2}),
\bfpage{857}
(\byear{2004}).
\arxivurl{astro-ph/0403678}.
doi:\doiurl{10.1086/424989}
\end{barticle}
\endbibitem

\bibitem[\protect\citeauthoryear{{Beckwith} et~al.}{2008}]{Beckwith2008}
\begin{barticle}
\bauthor{\bsnm{{Beckwith}}, \binits{K.}},
\bauthor{\bsnm{{Hawley}}, \binits{J.F.}},
\bauthor{\bsnm{{Krolik}}, \binits{J.H.}}:
\bjtitle{\apj}
\bvolume{678}(\bissue{2}),
\bfpage{1180}
(\byear{2008}).
\arxivurl{0709.3833}.
doi:\doiurl{10.1086/533492}
\end{barticle}
\endbibitem

\bibitem[\protect\citeauthoryear{{B{\'e}thune} et~al.}{2017}]{Bethune2017}
\begin{barticle}
\bauthor{\bsnm{{B{\'e}thune}}, \binits{W.}},
\bauthor{\bsnm{{Lesur}}, \binits{G.}},
\bauthor{\bsnm{{Ferreira}}, \binits{J.}}:
\bjtitle{\aap}
\bvolume{600},
\bfpage{75}
(\byear{2017}).
\arxivurl{1612.00883}.
doi:\doiurl{10.1051/0004-6361/201630056}
\end{barticle}
\endbibitem

\bibitem[\protect\citeauthoryear{{Bisnovatyi-Kogan} and
  {Lovelace}}{1997}]{Bisnovatyi1997}
\begin{barticle}
\bauthor{\bsnm{{Bisnovatyi-Kogan}}, \binits{G.S.}},
\bauthor{\bsnm{{Lovelace}}, \binits{R.V.E.}}:
\bjtitle{\apjl}
\bvolume{486}(\bissue{1}),
\bfpage{43}
(\byear{1997}).
\arxivurl{astro-ph/9704208}.
doi:\doiurl{10.1086/310826}
\end{barticle}
\endbibitem

\bibitem[\protect\citeauthoryear{{Bisnovatyi-Kogan} and
  {Ruzmaikin}}{1976}]{Bisnovatyi1976}
\begin{barticle}
\bauthor{\bsnm{{Bisnovatyi-Kogan}}, \binits{G.S.}},
\bauthor{\bsnm{{Ruzmaikin}}, \binits{A.A.}}:
\bjtitle{\apss}
\bvolume{42}(\bissue{2}),
\bfpage{401}
(\byear{1976}).
doi:\doiurl{10.1007/BF01225967}
\end{barticle}
\endbibitem

\bibitem[\protect\citeauthoryear{{Braginskii}}{1965}]{Braginskii1965}
\begin{barticle}
\bauthor{\bsnm{{Braginskii}}, \binits{S.I.}}:
\bjtitle{Reviews of Plasma Physics}
\bvolume{1},
\bfpage{205}
(\byear{1965})
\end{barticle}
\endbibitem

\bibitem[\protect\citeauthoryear{{Braiding} and {Wardle}}{2012}]{Braiding2012}
\begin{barticle}
\bauthor{\bsnm{{Braiding}}, \binits{C.R.}},
\bauthor{\bsnm{{Wardle}}, \binits{M.}}:
\bjtitle{\mnras}
\bvolume{422}(\bissue{1}),
\bfpage{261}
(\byear{2012}).
\arxivurl{1109.1370}.
doi:\doiurl{10.1111/j.1365-2966.2012.20601.x}
\end{barticle}
\endbibitem

\bibitem[\protect\citeauthoryear{{Bu} and {Yang}}{2019}]{Bu2019ApJ}
\begin{barticle}
\bauthor{\bsnm{{Bu}}, \binits{D.-F.}},
\bauthor{\bsnm{{Yang}}, \binits{X.-H.}}:
\bjtitle{\apj}
\bvolume{871}(\bissue{2}),
\bfpage{138}
(\byear{2019}).
doi:\doiurl{10.3847/1538-4357/aaf807}
\end{barticle}
\endbibitem

\bibitem[\protect\citeauthoryear{{Bu} et~al.}{2016}]{Bu2016MNRAS}
\begin{barticle}
\bauthor{\bsnm{{Bu}}, \binits{D.-F.}},
\bauthor{\bsnm{{Wu}}, \binits{M.-C.}},
\bauthor{\bsnm{{Yuan}}, \binits{Y.-F.}}:
\bjtitle{\mnras}
\bvolume{459}(\bissue{1}),
\bfpage{746}
(\byear{2016}).
\arxivurl{1603.07407}.
doi:\doiurl{10.1093/mnras/stw723}
\end{barticle}
\endbibitem

\bibitem[\protect\citeauthoryear{{Bu} et~al.}{2019}]{Bu2019Uni}
\begin{barticle}
\bauthor{\bsnm{{Bu}}, \binits{D.-F.}},
\bauthor{\bsnm{{Xu}}, \binits{P.-Y.}},
\bauthor{\bsnm{{Zhu}}, \binits{B.-C.}}:
\bjtitle{Universe}
\bvolume{5}(\bissue{4}),
\bfpage{89}
(\byear{2019}).
doi:\doiurl{10.3390/universe5040089}
\end{barticle}
\endbibitem

\bibitem[\protect\citeauthoryear{{Bu} et~al.}{2011}]{Bu2011}
\begin{barticle}
\bauthor{\bsnm{{Bu}}, \binits{D.-F.}},
\bauthor{\bsnm{{Yuan}}, \binits{F.}},
\bauthor{\bsnm{{Stone}}, \binits{J.M.}}:
\bjtitle{\mnras}
\bvolume{413}(\bissue{4}),
\bfpage{2808}
(\byear{2011}).
\arxivurl{1011.5331}.
doi:\doiurl{10.1111/j.1365-2966.2011.18354.x}
\end{barticle}
\endbibitem

\bibitem[\protect\citeauthoryear{{Bu} et~al.}{2009}]{Bu2009}
\begin{barticle}
\bauthor{\bsnm{{Bu}}, \binits{D.-F.}},
\bauthor{\bsnm{{Yuan}}, \binits{F.}},
\bauthor{\bsnm{{Xie}}, \binits{F.-G.}}:
\bjtitle{\mnras}
\bvolume{392}(\bissue{1}),
\bfpage{325}
(\byear{2009}).
\arxivurl{0810.1341}.
doi:\doiurl{10.1111/j.1365-2966.2008.14047.x}
\end{barticle}
\endbibitem

\bibitem[\protect\citeauthoryear{{Bu} et~al.}{2016a}]{Bu2016a}
\begin{barticle}
\bauthor{\bsnm{{Bu}}, \binits{D.-F.}},
\bauthor{\bsnm{{Yuan}}, \binits{F.}},
\bauthor{\bsnm{{Gan}}, \binits{Z.-M.}},
\bauthor{\bsnm{{Yang}}, \binits{X.-H.}}:
\bjtitle{\apj}
\bvolume{818}(\bissue{1}),
\bfpage{83}
(\byear{2016}a).
\arxivurl{1510.03124}.
doi:\doiurl{10.3847/0004-637X/818/1/83}
\end{barticle}
\endbibitem

\bibitem[\protect\citeauthoryear{{Bu} et~al.}{2016b}]{Bu2016b}
\begin{barticle}
\bauthor{\bsnm{{Bu}}, \binits{D.-F.}},
\bauthor{\bsnm{{Yuan}}, \binits{F.}},
\bauthor{\bsnm{{Gan}}, \binits{Z.-M.}},
\bauthor{\bsnm{{Yang}}, \binits{X.-H.}}:
\bjtitle{\apj}
\bvolume{823}(\bissue{2}),
\bfpage{90}
(\byear{2016}b).
\arxivurl{1603.09442}.
doi:\doiurl{10.3847/0004-637X/823/2/90}
\end{barticle}
\endbibitem

\bibitem[\protect\citeauthoryear{{Cao}}{2016}]{Cao2016}
\begin{barticle}
\bauthor{\bsnm{{Cao}}, \binits{X.}}:
\bjtitle{\apj}
\bvolume{817}(\bissue{1}),
\bfpage{71}
(\byear{2016}).
\arxivurl{1512.00124}.
doi:\doiurl{10.3847/0004-637X/817/1/71}
\end{barticle}
\endbibitem

\bibitem[\protect\citeauthoryear{{Cao} and {Lai}}{2019}]{Cao2019}
\begin{barticle}
\bauthor{\bsnm{{Cao}}, \binits{X.}},
\bauthor{\bsnm{{Lai}}, \binits{D.}}:
\bjtitle{\mnras}
\bvolume{485}(\bissue{2}),
\bfpage{1916}
(\byear{2019}).
\arxivurl{1712.09265}.
doi:\doiurl{10.1093/mnras/stz580}
\end{barticle}
\endbibitem

\bibitem[\protect\citeauthoryear{{Chandra} et~al.}{2015}]{Chandra2015}
\begin{barticle}
\bauthor{\bsnm{{Chandra}}, \binits{M.}},
\bauthor{\bsnm{{Gammie}}, \binits{C.F.}},
\bauthor{\bsnm{{Foucart}}, \binits{F.}},
\bauthor{\bsnm{{Quataert}}, \binits{E.}}:
\bjtitle{\apj}
\bvolume{810}(\bissue{2}),
\bfpage{162}
(\byear{2015}).
\arxivurl{1508.00878}.
doi:\doiurl{10.1088/0004-637X/810/2/162}
\end{barticle}
\endbibitem

\bibitem[\protect\citeauthoryear{{Cheung} et~al.}{2016}]{Cheung2016}
\begin{barticle}
\bauthor{\bsnm{{Cheung}}, \binits{E.}},
\bauthor{\bsnm{{Bundy}}, \binits{K.}},
\bauthor{\bsnm{{Cappellari}}, \binits{M.}},
\bauthor{\bsnm{{Peirani}}, \binits{S.}},
\bauthor{\bsnm{{Rujopakarn}}, \binits{W.}},
\bauthor{\bsnm{{Westfall}}, \binits{K.}},
\bauthor{\bsnm{{Yan}}, \binits{R.}},
\bauthor{\bsnm{{Bershady}}, \binits{M.}},
\bauthor{\bsnm{{Greene}}, \binits{J.E.}},
\bauthor{\bsnm{{Heckman}}, \binits{T.M.}},
\bauthor{\bsnm{{Drory}}, \binits{N.}},
\bauthor{\bsnm{{Law}}, \binits{D.R.}},
\bauthor{\bsnm{{Masters}}, \binits{K.L.}},
\bauthor{\bsnm{{Thomas}}, \binits{D.}},
\bauthor{\bsnm{{Wake}}, \binits{D.A.}},
\bauthor{\bsnm{{Weijmans}}, \binits{A.-M.}},
\bauthor{\bsnm{{Rubin}}, \binits{K.}},
\bauthor{\bsnm{{Belfiore}}, \binits{F.}},
\bauthor{\bsnm{{Vulcani}}, \binits{B.}},
\bauthor{\bsnm{{Chen}}, \binits{Y.-M.}},
\bauthor{\bsnm{{Zhang}}, \binits{K.}},
\bauthor{\bsnm{{Gelfand}}, \binits{J.D.}},
\bauthor{\bsnm{{Bizyaev}}, \binits{D.}},
\bauthor{\bsnm{{Roman-Lopes}}, \binits{A.}},
\bauthor{\bsnm{{Schneider}}, \binits{D.P.}}:
\bjtitle{\nat}
\bvolume{533}(\bissue{7604}),
\bfpage{504}
(\byear{2016}).
\arxivurl{1605.07626}.
doi:\doiurl{10.1038/nature18006}
\end{barticle}
\endbibitem

\bibitem[\protect\citeauthoryear{{Crenshaw} and {Kraemer}}{2012}]{Crenshaw2012}
\begin{barticle}
\bauthor{\bsnm{{Crenshaw}}, \binits{D.M.}},
\bauthor{\bsnm{{Kraemer}}, \binits{S.B.}}:
\bjtitle{\apj}
\bvolume{753}(\bissue{1}),
\bfpage{75}
(\byear{2012}).
\arxivurl{1204.6694}.
doi:\doiurl{10.1088/0004-637X/753/1/75}
\end{barticle}
\endbibitem

\bibitem[\protect\citeauthoryear{{Das} and {Basu}}{2021}]{Das2021}
\begin{barticle}
\bauthor{\bsnm{{Das}}, \binits{I.}},
\bauthor{\bsnm{{Basu}}, \binits{S.}}:
\bjtitle{\apj}
\bvolume{910}(\bissue{2}),
\bfpage{163}
(\byear{2021}).
\arxivurl{2011.08876}.
doi:\doiurl{10.3847/1538-4357/abdb2c}
\end{barticle}
\endbibitem

\bibitem[\protect\citeauthoryear{{De Villiers} et~al.}{2003}]{DeVilliers2003}
\begin{barticle}
\bauthor{\bsnm{{De Villiers}}, \binits{J.-P.}},
\bauthor{\bsnm{{Hawley}}, \binits{J.F.}},
\bauthor{\bsnm{{Krolik}}, \binits{J.H.}}:
\bjtitle{\apj}
\bvolume{599}(\bissue{2}),
\bfpage{1238}
(\byear{2003}).
\arxivurl{astro-ph/0307260}.
doi:\doiurl{10.1086/379509}
\end{barticle}
\endbibitem

\bibitem[\protect\citeauthoryear{{Ding} et~al.}{2010}]{Ding2010}
\begin{barticle}
\bauthor{\bsnm{{Ding}}, \binits{J.}},
\bauthor{\bsnm{{Yuan}}, \binits{F.}},
\bauthor{\bsnm{{Liang}}, \binits{E.}}:
\bjtitle{\apj}
\bvolume{708}(\bissue{2}),
\bfpage{1545}
(\byear{2010}).
\arxivurl{0911.4560}.
doi:\doiurl{10.1088/0004-637X/708/2/1545}
\end{barticle}
\endbibitem

\bibitem[\protect\citeauthoryear{{Done} et~al.}{2007}]{Done2007}
\begin{barticle}
\bauthor{\bsnm{{Done}}, \binits{C.}},
\bauthor{\bsnm{{Gierli{\'n}ski}}, \binits{M.}},
\bauthor{\bsnm{{Kubota}}, \binits{A.}}:
\bjtitle{\aapr}
\bvolume{15}(\bissue{1}),
\bfpage{1}
(\byear{2007}).
\arxivurl{0708.0148}.
doi:\doiurl{10.1007/s00159-007-0006-1}
\end{barticle}
\endbibitem

\bibitem[\protect\citeauthoryear{{Fendt} and
  {{\v{C}}emelji{\'c}}}{2002}]{Fendt2002}
\begin{barticle}
\bauthor{\bsnm{{Fendt}}, \binits{C.}},
\bauthor{\bsnm{{{\v{C}}emelji{\'c}}}, \binits{M.}}:
\bjtitle{\aap}
\bvolume{395},
\bfpage{1045}
(\byear{2002}).
\arxivurl{astro-ph/0210082}.
doi:\doiurl{10.1051/0004-6361:20021442}
\end{barticle}
\endbibitem

\bibitem[\protect\citeauthoryear{{Fleming} et~al.}{2000}]{Fleming2000}
\begin{barticle}
\bauthor{\bsnm{{Fleming}}, \binits{T.P.}},
\bauthor{\bsnm{{Stone}}, \binits{J.M.}},
\bauthor{\bsnm{{Hawley}}, \binits{J.F.}}:
\bjtitle{\apj}
\bvolume{530}(\bissue{1}),
\bfpage{464}
(\byear{2000}).
\arxivurl{astro-ph/0001164}.
doi:\doiurl{10.1086/308338}
\end{barticle}
\endbibitem

\bibitem[\protect\citeauthoryear{{Ghoreyshi}}{2020}]{Ghoreyshi2020PASA}
\begin{barticle}
\bauthor{\bsnm{{Ghoreyshi}}, \binits{S.M.}}:
\bjtitle{\pasa}
\bvolume{37},
\bfpage{023}
(\byear{2020}).
\arxivurl{2004.14757}.
doi:\doiurl{10.1017/pasa.2020.14}
\end{barticle}
\endbibitem

\bibitem[\protect\citeauthoryear{{Ghoreyshi} and
  {Shadmehri}}{2020}]{Ghoreyshi2020}
\begin{botherref}
\oauthor{\bsnm{{Ghoreyshi}}, \binits{S.M.}},
\oauthor{\bsnm{{Shadmehri}}, \binits{M.}}:
\mnras
(2020).
\arxivurl{2003.04752}.
doi:\doiurl{10.1093/mnras/staa599}
\end{botherref}
\endbibitem

\bibitem[\protect\citeauthoryear{{Gravity Collaboration}
  et~al.}{2018}]{Gravity2018}
\begin{barticle}
\bauthor{\bsnm{{Gravity Collaboration}}},
\bauthor{\bsnm{{Abuter}}, \binits{R.}},
\bauthor{\bsnm{{Amorim}}, \binits{A.}},
\bauthor{\bsnm{{Baub{\"o}ck}}, \binits{M.}},
\bauthor{\bsnm{{Berger}}, \binits{J.P.}},
\bauthor{\bsnm{{Bonnet}}, \binits{H.}},
\bauthor{\bsnm{{Brandner}}, \binits{W.}},
\bauthor{\bsnm{{Cl{\'e}net}}, \binits{Y.}},
\bauthor{\bsnm{{Coud{\'e} Du Foresto}}, \binits{V.}},
\bauthor{\bsnm{{de Zeeuw}}, \binits{P.T.}},
\bauthor{\bsnm{{Deen}}, \binits{C.}},
\bauthor{\bsnm{{Dexter}}, \binits{J.}},
\bauthor{\bsnm{{Duvert}}, \binits{G.}},
\bauthor{\bsnm{{Eckart}}, \binits{A.}},
\bauthor{\bsnm{{Eisenhauer}}, \binits{F.}},
\bauthor{\bsnm{{F{\"o}rster Schreiber}}, \binits{N.M.}},
\bauthor{\bsnm{{Garcia}}, \binits{P.}},
\bauthor{\bsnm{{Gao}}, \binits{F.}},
\bauthor{\bsnm{{Gendron}}, \binits{E.}},
\bauthor{\bsnm{{Genzel}}, \binits{R.}},
\bauthor{\bsnm{{Gillessen}}, \binits{S.}},
\bauthor{\bsnm{{Guajardo}}, \binits{P.}},
\bauthor{\bsnm{{Habibi}}, \binits{M.}},
\bauthor{\bsnm{{Haubois}}, \binits{X.}},
\bauthor{\bsnm{{Henning}}, \binits{T.}},
\bauthor{\bsnm{{Hippler}}, \binits{S.}},
\bauthor{\bsnm{{Horrobin}}, \binits{M.}},
\bauthor{\bsnm{{Huber}}, \binits{A.}},
\bauthor{\bsnm{{Jim{\'e}nez-Rosales}}, \binits{A.}},
\bauthor{\bsnm{{Jocou}}, \binits{L.}},
\bauthor{\bsnm{{Kervella}}, \binits{P.}},
\bauthor{\bsnm{{Lacour}}, \binits{S.}},
\bauthor{\bsnm{{Lapeyr{\`e}re}}, \binits{V.}},
\bauthor{\bsnm{{Lazareff}}, \binits{B.}},
\bauthor{\bsnm{{Le Bouquin}}, \binits{J.-B.}},
\bauthor{\bsnm{{L{\'e}na}}, \binits{P.}},
\bauthor{\bsnm{{Lippa}}, \binits{M.}},
\bauthor{\bsnm{{Ott}}, \binits{T.}},
\bauthor{\bsnm{{Panduro}}, \binits{J.}},
\bauthor{\bsnm{{Paumard}}, \binits{T.}},
\bauthor{\bsnm{{Perraut}}, \binits{K.}},
\bauthor{\bsnm{{Perrin}}, \binits{G.}},
\bauthor{\bsnm{{Pfuhl}}, \binits{O.}},
\bauthor{\bsnm{{Plewa}}, \binits{P.M.}},
\bauthor{\bsnm{{Rabien}}, \binits{S.}},
\bauthor{\bsnm{{Rodr{\'\i}guez-Coira}}, \binits{G.}},
\bauthor{\bsnm{{Rousset}}, \binits{G.}},
\bauthor{\bsnm{{Sternberg}}, \binits{A.}},
\bauthor{\bsnm{{Straub}}, \binits{O.}},
\bauthor{\bsnm{{Straubmeier}}, \binits{C.}},
\bauthor{\bsnm{{Sturm}}, \binits{E.}},
\bauthor{\bsnm{{Tacconi}}, \binits{L.J.}},
\bauthor{\bsnm{{Vincent}}, \binits{F.}},
\bauthor{\bsnm{{von Fellenberg}}, \binits{S.}},
\bauthor{\bsnm{{Waisberg}}, \binits{I.}},
\bauthor{\bsnm{{Widmann}}, \binits{F.}},
\bauthor{\bsnm{{Wieprecht}}, \binits{E.}},
\bauthor{\bsnm{{Wiezorrek}}, \binits{E.}},
\bauthor{\bsnm{{Woillez}}, \binits{J.}},
\bauthor{\bsnm{{Yazici}}, \binits{S.}}:
\bjtitle{\aap}
\bvolume{618},
\bfpage{10}
(\byear{2018}).
\arxivurl{1810.12641}.
doi:\doiurl{10.1051/0004-6361/201834294}
\end{barticle}
\endbibitem

\bibitem[\protect\citeauthoryear{{Hirose} et~al.}{2004}]{Hirose2004}
\begin{barticle}
\bauthor{\bsnm{{Hirose}}, \binits{S.}},
\bauthor{\bsnm{{Krolik}}, \binits{J.H.}},
\bauthor{\bsnm{{De Villiers}}, \binits{J.-P.}},
\bauthor{\bsnm{{Hawley}}, \binits{J.F.}}:
\bjtitle{\apj}
\bvolume{606}(\bissue{2}),
\bfpage{1083}
(\byear{2004}).
\arxivurl{astro-ph/0311500}.
doi:\doiurl{10.1086/383184}
\end{barticle}
\endbibitem

\bibitem[\protect\citeauthoryear{{Homan} et~al.}{2016}]{Homan2016}
\begin{barticle}
\bauthor{\bsnm{{Homan}}, \binits{J.}},
\bauthor{\bsnm{{Neilsen}}, \binits{J.}},
\bauthor{\bsnm{{Allen}}, \binits{J.L.}},
\bauthor{\bsnm{{Chakrabarty}}, \binits{D.}},
\bauthor{\bsnm{{Fender}}, \binits{R.}},
\bauthor{\bsnm{{Fridriksson}}, \binits{J.K.}},
\bauthor{\bsnm{{Remillard}}, \binits{R.A.}},
\bauthor{\bsnm{{Schulz}}, \binits{N.}}:
\bjtitle{\apjl}
\bvolume{830}(\bissue{1}),
\bfpage{5}
(\byear{2016}).
\arxivurl{1606.07954}.
doi:\doiurl{10.3847/2041-8205/830/1/L5}
\end{barticle}
\endbibitem

\bibitem[\protect\citeauthoryear{{Igumenshchev}
  et~al.}{2003}]{Igumenshchev2003}
\begin{barticle}
\bauthor{\bsnm{{Igumenshchev}}, \binits{I.V.}},
\bauthor{\bsnm{{Narayan}}, \binits{R.}},
\bauthor{\bsnm{{Abramowicz}}, \binits{M.A.}}:
\bjtitle{\apj}
\bvolume{592}(\bissue{2}),
\bfpage{1042}
(\byear{2003}).
\arxivurl{astro-ph/0301402}.
doi:\doiurl{10.1086/375769}
\end{barticle}
\endbibitem

\bibitem[\protect\citeauthoryear{{Kato} et~al.}{2008}]{Kato2008}
\begin{bbook}
\bauthor{\bsnm{{Kato}}, \binits{S.}},
\bauthor{\bsnm{{Fukue}}, \binits{J.}},
\bauthor{\bsnm{{Mineshige}}, \binits{S.}}:
\bbtitle{{Black-Hole Accretion Disks --- Towards a New Paradigm ---}},
(\byear{2008})
\end{bbook}
\endbibitem

\bibitem[\protect\citeauthoryear{{Kempski} et~al.}{2019}]{Kempski2019}
\begin{barticle}
\bauthor{\bsnm{{Kempski}}, \binits{P.}},
\bauthor{\bsnm{{Quataert}}, \binits{E.}},
\bauthor{\bsnm{{Squire}}, \binits{J.}},
\bauthor{\bsnm{{Kunz}}, \binits{M.W.}}:
\bjtitle{\mnras}
\bvolume{486}(\bissue{3}),
\bfpage{4013}
(\byear{2019}).
\arxivurl{1901.04504}.
doi:\doiurl{10.1093/mnras/stz1111}
\end{barticle}
\endbibitem

\bibitem[\protect\citeauthoryear{{Knigge}}{1999}]{Knigge1999}
\begin{barticle}
\bauthor{\bsnm{{Knigge}}, \binits{C.}}:
\bjtitle{\mnras}
\bvolume{309},
\bfpage{409}
(\byear{1999}).
\arxivurl{astro-ph/9906194}.
doi:\doiurl{10.1046/j.1365-8711.1999.02839.x}
\end{barticle}
\endbibitem

\bibitem[\protect\citeauthoryear{{Kuwabara} et~al.}{2000}]{Kuwabara2000}
\begin{barticle}
\bauthor{\bsnm{{Kuwabara}}, \binits{T.}},
\bauthor{\bsnm{{Shibata}}, \binits{K.}},
\bauthor{\bsnm{{Kudoh}}, \binits{T.}},
\bauthor{\bsnm{{Matsumoto}}, \binits{R.}}:
\bjtitle{\pasj}
\bvolume{52},
\bfpage{1109}
(\byear{2000}).
\arxivurl{astro-ph/0011165}.
doi:\doiurl{10.1093/pasj/52.6.1109}
\end{barticle}
\endbibitem

\bibitem[\protect\citeauthoryear{{Ma} et~al.}{2019}]{Ma2019}
\begin{barticle}
\bauthor{\bsnm{{Ma}}, \binits{R.-Y.}},
\bauthor{\bsnm{{Roberts}}, \binits{S.R.}},
\bauthor{\bsnm{{Li}}, \binits{Y.-P.}},
\bauthor{\bsnm{{Wang}}, \binits{Q.D.}}:
\bjtitle{\mnras}
\bvolume{483}(\bissue{4}),
\bfpage{5614}
(\byear{2019}).
\arxivurl{1811.02190}.
doi:\doiurl{10.1093/mnras/sty3039}
\end{barticle}
\endbibitem

\bibitem[\protect\citeauthoryear{{Mori} et~al.}{2019}]{Mori2019}
\begin{barticle}
\bauthor{\bsnm{{Mori}}, \binits{S.}},
\bauthor{\bsnm{{Bai}}, \binits{X.-N.}},
\bauthor{\bsnm{{Okuzumi}}, \binits{S.}}:
\bjtitle{\apj}
\bvolume{872}(\bissue{1}),
\bfpage{98}
(\byear{2019}).
\arxivurl{1901.06921}.
doi:\doiurl{10.3847/1538-4357/ab0022}
\end{barticle}
\endbibitem

\bibitem[\protect\citeauthoryear{{Mosallanezhad}
  et~al.}{2013}]{Mosallanezhad2013}
\begin{barticle}
\bauthor{\bsnm{{Mosallanezhad}}, \binits{A.}},
\bauthor{\bsnm{{Khajavi}}, \binits{M.}},
\bauthor{\bsnm{{Abbassi}}, \binits{S.}}:
\bjtitle{Research in Astronomy and Astrophysics}
\bvolume{13}(\bissue{1}),
\bfpage{87}
(\byear{2013}).
\arxivurl{1206.2841}.
doi:\doiurl{10.1088/1674-4527/13/1/009}
\end{barticle}
\endbibitem

\bibitem[\protect\citeauthoryear{{Narayan} and {Yi}}{1994}]{Narayan1994}
\begin{barticle}
\bauthor{\bsnm{{Narayan}}, \binits{R.}},
\bauthor{\bsnm{{Yi}}, \binits{I.}}:
\bjtitle{\apjl}
\bvolume{428},
\bfpage{13}
(\byear{1994}).
\arxivurl{astro-ph/9403052}.
doi:\doiurl{10.1086/187381}
\end{barticle}
\endbibitem

\bibitem[\protect\citeauthoryear{{Ohsuga} et~al.}{2009}]{Ohsuga2009}
\begin{barticle}
\bauthor{\bsnm{{Ohsuga}}, \binits{K.}},
\bauthor{\bsnm{{Mineshige}}, \binits{S.}},
\bauthor{\bsnm{{Mori}}, \binits{M.}},
\bauthor{\bsnm{{Kato}}, \binits{Y.}}:
\bjtitle{\pasj}
\bvolume{61}(\bissue{3}),
\bfpage{7}
(\byear{2009}).
\arxivurl{0903.5364}.
doi:\doiurl{10.1093/pasj/61.3.L7}
\end{barticle}
\endbibitem

\bibitem[\protect\citeauthoryear{{Pandey} and {Wardle}}{2008}]{Pandey2008}
\begin{barticle}
\bauthor{\bsnm{{Pandey}}, \binits{B.P.}},
\bauthor{\bsnm{{Wardle}}, \binits{M.}}:
\bjtitle{\mnras}
\bvolume{385}(\bissue{4}),
\bfpage{2269}
(\byear{2008}).
\arxivurl{0707.2688}.
doi:\doiurl{10.1111/j.1365-2966.2008.12998.x}
\end{barticle}
\endbibitem

\bibitem[\protect\citeauthoryear{{Park} et~al.}{2019}]{Park2019}
\begin{barticle}
\bauthor{\bsnm{{Park}}, \binits{J.}},
\bauthor{\bsnm{{Hada}}, \binits{K.}},
\bauthor{\bsnm{{Kino}}, \binits{M.}},
\bauthor{\bsnm{{Nakamura}}, \binits{M.}},
\bauthor{\bsnm{{Ro}}, \binits{H.}},
\bauthor{\bsnm{{Trippe}}, \binits{S.}}:
\bjtitle{\apj}
\bvolume{871}(\bissue{2}),
\bfpage{257}
(\byear{2019}).
\arxivurl{1812.08386}.
doi:\doiurl{10.3847/1538-4357/aaf9a9}
\end{barticle}
\endbibitem

\bibitem[\protect\citeauthoryear{{Parrish} and {Stone}}{2007}]{Parrish2007}
\begin{barticle}
\bauthor{\bsnm{{Parrish}}, \binits{I.J.}},
\bauthor{\bsnm{{Stone}}, \binits{J.M.}}:
\bjtitle{\apj}
\bvolume{664}(\bissue{1}),
\bfpage{135}
(\byear{2007}).
\arxivurl{astro-ph/0612195}.
doi:\doiurl{10.1086/518881}
\end{barticle}
\endbibitem

\bibitem[\protect\citeauthoryear{{Pudritz}}{1985}]{Pudritz1985}
\begin{barticle}
\bauthor{\bsnm{{Pudritz}}, \binits{R.E.}}:
\bjtitle{\apj}
\bvolume{293},
\bfpage{216}
(\byear{1985}).
doi:\doiurl{10.1086/163227}
\end{barticle}
\endbibitem

\bibitem[\protect\citeauthoryear{{Qian} et~al.}{2018}]{Qian2018}
\begin{barticle}
\bauthor{\bsnm{{Qian}}, \binits{Q.}},
\bauthor{\bsnm{{Fendt}}, \binits{C.}},
\bauthor{\bsnm{{Vourellis}}, \binits{C.}}:
\bjtitle{\apj}
\bvolume{859}(\bissue{1}),
\bfpage{28}
(\byear{2018}).
\arxivurl{1804.09652}.
doi:\doiurl{10.3847/1538-4357/aabd36}
\end{barticle}
\endbibitem

\bibitem[\protect\citeauthoryear{{Qiao} and {Liu}}{2009}]{Qiao2009}
\begin{barticle}
\bauthor{\bsnm{{Qiao}}, \binits{E.}},
\bauthor{\bsnm{{Liu}}, \binits{B.F.}}:
\bjtitle{\pasj}
\bvolume{61},
\bfpage{403}
(\byear{2009}).
\arxivurl{0901.0475}.
doi:\doiurl{10.1093/pasj/61.2.403}
\end{barticle}
\endbibitem

\bibitem[\protect\citeauthoryear{{Quataert}}{1998}]{Quataert1998}
\begin{barticle}
\bauthor{\bsnm{{Quataert}}, \binits{E.}}:
\bjtitle{\apj}
\bvolume{500}(\bissue{2}),
\bfpage{978}
(\byear{1998}).
\arxivurl{astro-ph/9710127}.
doi:\doiurl{10.1086/305770}
\end{barticle}
\endbibitem

\bibitem[\protect\citeauthoryear{{Quataert} et~al.}{2002}]{Quataert2002}
\begin{barticle}
\bauthor{\bsnm{{Quataert}}, \binits{E.}},
\bauthor{\bsnm{{Dorland}}, \binits{W.}},
\bauthor{\bsnm{{Hammett}}, \binits{G.W.}}:
\bjtitle{\apj}
\bvolume{577}(\bissue{1}),
\bfpage{524}
(\byear{2002}).
\arxivurl{astro-ph/0205492}.
doi:\doiurl{10.1086/342174}
\end{barticle}
\endbibitem

\bibitem[\protect\citeauthoryear{{Ripperda} et~al.}{2019a}]{Ripperda2019ApJS}
\begin{barticle}
\bauthor{\bsnm{{Ripperda}}, \binits{B.}},
\bauthor{\bsnm{{Bacchini}}, \binits{F.}},
\bauthor{\bsnm{{Porth}}, \binits{O.}},
\bauthor{\bsnm{{Most}}, \binits{E.R.}},
\bauthor{\bsnm{{Olivares}}, \binits{H.}},
\bauthor{\bsnm{{Nathanail}}, \binits{A.}},
\bauthor{\bsnm{{Rezzolla}}, \binits{L.}},
\bauthor{\bsnm{{Teunissen}}, \binits{J.}},
\bauthor{\bsnm{{Keppens}}, \binits{R.}}:
\bjtitle{\apjs}
\bvolume{244}(\bissue{1}),
\bfpage{10}
(\byear{2019}a).
\arxivurl{1907.07197}.
doi:\doiurl{10.3847/1538-4365/ab3922}
\end{barticle}
\endbibitem

\bibitem[\protect\citeauthoryear{{Ripperda} et~al.}{2019b}]{Ripperda2019mnras}
\begin{barticle}
\bauthor{\bsnm{{Ripperda}}, \binits{B.}},
\bauthor{\bsnm{{Porth}}, \binits{O.}},
\bauthor{\bsnm{{Sironi}}, \binits{L.}},
\bauthor{\bsnm{{Keppens}}, \binits{R.}}:
\bjtitle{\mnras}
\bvolume{485}(\bissue{1}),
\bfpage{299}
(\byear{2019}b).
\arxivurl{1810.10116}.
doi:\doiurl{10.1093/mnras/stz387}
\end{barticle}
\endbibitem

\bibitem[\protect\citeauthoryear{{Ripperda} et~al.}{2020}]{Ripperda2020}
\begin{barticle}
\bauthor{\bsnm{{Ripperda}}, \binits{B.}},
\bauthor{\bsnm{{Bacchini}}, \binits{F.}},
\bauthor{\bsnm{{Philippov}}, \binits{A.A.}}:
\bjtitle{\apj}
\bvolume{900}(\bissue{2}),
\bfpage{100}
(\byear{2020}).
\arxivurl{2003.04330}.
doi:\doiurl{10.3847/1538-4357/ababab}
\end{barticle}
\endbibitem

\bibitem[\protect\citeauthoryear{{Riquelme} et~al.}{2012}]{Riquelme2012}
\begin{barticle}
\bauthor{\bsnm{{Riquelme}}, \binits{M.A.}},
\bauthor{\bsnm{{Quataert}}, \binits{E.}},
\bauthor{\bsnm{{Sharma}}, \binits{P.}},
\bauthor{\bsnm{{Spitkovsky}}, \binits{A.}}:
\bjtitle{\apj}
\bvolume{755}(\bissue{1}),
\bfpage{50}
(\byear{2012}).
\arxivurl{1201.6407}.
doi:\doiurl{10.1088/0004-637X/755/1/50}
\end{barticle}
\endbibitem

\bibitem[\protect\citeauthoryear{{Samadi} et~al.}{2014}]{Samadi2014}
\begin{barticle}
\bauthor{\bsnm{{Samadi}}, \binits{M.}},
\bauthor{\bsnm{{Abbassi}}, \binits{S.}},
\bauthor{\bsnm{{Khajavi}}, \binits{M.}}:
\bjtitle{\mnras}
\bvolume{437}(\bissue{4}),
\bfpage{3124}
(\byear{2014}).
\arxivurl{1310.6317}.
doi:\doiurl{10.1093/mnras/stt2052}
\end{barticle}
\endbibitem

\bibitem[\protect\citeauthoryear{{Sano} and {Stone}}{2002}]{Sano2002}
\begin{barticle}
\bauthor{\bsnm{{Sano}}, \binits{T.}},
\bauthor{\bsnm{{Stone}}, \binits{J.M.}}:
\bjtitle{\apj}
\bvolume{570}(\bissue{1}),
\bfpage{314}
(\byear{2002}).
\arxivurl{astro-ph/0201179}.
doi:\doiurl{10.1086/339504}
\end{barticle}
\endbibitem

\bibitem[\protect\citeauthoryear{{Scepi} et~al.}{2018}]{Scepi2018}
\begin{barticle}
\bauthor{\bsnm{{Scepi}}, \binits{N.}},
\bauthor{\bsnm{{Lesur}}, \binits{G.}},
\bauthor{\bsnm{{Dubus}}, \binits{G.}},
\bauthor{\bsnm{{Flock}}, \binits{M.}}:
\bjtitle{\aap}
\bvolume{609},
\bfpage{77}
(\byear{2018}).
\arxivurl{1710.05872}.
doi:\doiurl{10.1051/0004-6361/201731900}
\end{barticle}
\endbibitem

\bibitem[\protect\citeauthoryear{{Shadmehri}}{2004}]{Shadmehri2004}
\begin{barticle}
\bauthor{\bsnm{{Shadmehri}}, \binits{M.}}:
\bjtitle{\aap}
\bvolume{424},
\bfpage{379}
(\byear{2004}).
\arxivurl{astro-ph/0303220}.
doi:\doiurl{10.1051/0004-6361:20040538}
\end{barticle}
\endbibitem

\bibitem[\protect\citeauthoryear{{Shakura} and {Sunyaev}}{1973}]{Shakura1973}
\begin{barticle}
\bauthor{\bsnm{{Shakura}}, \binits{N.I.}},
\bauthor{\bsnm{{Sunyaev}}, \binits{R.A.}}:
\bjtitle{\aap}
\bvolume{24},
\bfpage{337}
(\byear{1973})
\end{barticle}
\endbibitem

\bibitem[\protect\citeauthoryear{{Sharma} et~al.}{2008}]{Sharma2008}
\begin{barticle}
\bauthor{\bsnm{{Sharma}}, \binits{P.}},
\bauthor{\bsnm{{Quataert}}, \binits{E.}},
\bauthor{\bsnm{{Stone}}, \binits{J.M.}}:
\bjtitle{\mnras}
\bvolume{389}(\bissue{4}),
\bfpage{1815}
(\byear{2008}).
\arxivurl{0804.1353}.
doi:\doiurl{10.1111/j.1365-2966.2008.13686.x}
\end{barticle}
\endbibitem

\bibitem[\protect\citeauthoryear{{Sharma} et~al.}{2003}]{Sharma2003}
\begin{barticle}
\bauthor{\bsnm{{Sharma}}, \binits{P.}},
\bauthor{\bsnm{{Hammett}}, \binits{G.W.}},
\bauthor{\bsnm{{Quataert}}, \binits{E.}}:
\bjtitle{\apj}
\bvolume{596}(\bissue{2}),
\bfpage{1121}
(\byear{2003}).
\arxivurl{astro-ph/0305486}.
doi:\doiurl{10.1086/378234}
\end{barticle}
\endbibitem

\bibitem[\protect\citeauthoryear{{Sharma} et~al.}{2006}]{Sharma2006}
\begin{barticle}
\bauthor{\bsnm{{Sharma}}, \binits{P.}},
\bauthor{\bsnm{{Hammett}}, \binits{G.W.}},
\bauthor{\bsnm{{Quataert}}, \binits{E.}},
\bauthor{\bsnm{{Stone}}, \binits{J.M.}}:
\bjtitle{\apj}
\bvolume{637}(\bissue{2}),
\bfpage{952}
(\byear{2006}).
\arxivurl{astro-ph/0508502}.
doi:\doiurl{10.1086/498405}
\end{barticle}
\endbibitem

\bibitem[\protect\citeauthoryear{{Sharma} et~al.}{2007}]{Sharma2007}
\begin{barticle}
\bauthor{\bsnm{{Sharma}}, \binits{P.}},
\bauthor{\bsnm{{Quataert}}, \binits{E.}},
\bauthor{\bsnm{{Hammett}}, \binits{G.W.}},
\bauthor{\bsnm{{Stone}}, \binits{J.M.}}:
\bjtitle{\apj}
\bvolume{667}(\bissue{2}),
\bfpage{714}
(\byear{2007}).
\arxivurl{astro-ph/0703572}.
doi:\doiurl{10.1086/520800}
\end{barticle}
\endbibitem

\bibitem[\protect\citeauthoryear{{Tombesi} et~al.}{2010}]{Tombesi2010}
\begin{barticle}
\bauthor{\bsnm{{Tombesi}}, \binits{F.}},
\bauthor{\bsnm{{Sambruna}}, \binits{R.M.}},
\bauthor{\bsnm{{Reeves}}, \binits{J.N.}},
\bauthor{\bsnm{{Braito}}, \binits{V.}},
\bauthor{\bsnm{{Ballo}}, \binits{L.}},
\bauthor{\bsnm{{Gofford}}, \binits{J.}},
\bauthor{\bsnm{{Cappi}}, \binits{M.}},
\bauthor{\bsnm{{Mushotzky}}, \binits{R.F.}}:
\bjtitle{\apj}
\bvolume{719}(\bissue{1}),
\bfpage{700}
(\byear{2010}).
\arxivurl{1006.3536}.
doi:\doiurl{10.1088/0004-637X/719/1/700}
\end{barticle}
\endbibitem

\bibitem[\protect\citeauthoryear{{Tombesi} et~al.}{2014}]{Tombesi2014}
\begin{barticle}
\bauthor{\bsnm{{Tombesi}}, \binits{F.}},
\bauthor{\bsnm{{Tazaki}}, \binits{F.}},
\bauthor{\bsnm{{Mushotzky}}, \binits{R.F.}},
\bauthor{\bsnm{{Ueda}}, \binits{Y.}},
\bauthor{\bsnm{{Cappi}}, \binits{M.}},
\bauthor{\bsnm{{Gofford}}, \binits{J.}},
\bauthor{\bsnm{{Reeves}}, \binits{J.N.}},
\bauthor{\bsnm{{Guainazzi}}, \binits{M.}}:
\bjtitle{\mnras}
\bvolume{443}(\bissue{3}),
\bfpage{2154}
(\byear{2014}).
\arxivurl{1406.7252}.
doi:\doiurl{10.1093/mnras/stu1297}
\end{barticle}
\endbibitem

\bibitem[\protect\citeauthoryear{{{\v{C}}emelji{\'c}}
  et~al.}{2014}]{Cemeljic2014}
\begin{barticle}
\bauthor{\bsnm{{{\v{C}}emelji{\'c}}}, \binits{M.}},
\bauthor{\bsnm{{Vlahakis}}, \binits{N.}},
\bauthor{\bsnm{{Tsinganos}}, \binits{K.}}:
\bjtitle{\mnras}
\bvolume{442}(\bissue{2}),
\bfpage{1133}
(\byear{2014}).
\arxivurl{1405.3924}.
doi:\doiurl{10.1093/mnras/stu952}
\end{barticle}
\endbibitem

\bibitem[\protect\citeauthoryear{{Vourellis} et~al.}{2019}]{Vourellis2019}
\begin{barticle}
\bauthor{\bsnm{{Vourellis}}, \binits{C.}},
\bauthor{\bsnm{{Fendt}}, \binits{C.}},
\bauthor{\bsnm{{Qian}}, \binits{Q.}},
\bauthor{\bsnm{{Noble}}, \binits{S.C.}}:
\bjtitle{\apj}
\bvolume{882}(\bissue{1}),
\bfpage{2}
(\byear{2019}).
\arxivurl{1907.10622}.
doi:\doiurl{10.3847/1538-4357/ab32e2}
\end{barticle}
\endbibitem

\bibitem[\protect\citeauthoryear{{Wang} et~al.}{2019}]{Wang2019}
\begin{barticle}
\bauthor{\bsnm{{Wang}}, \binits{L.}},
\bauthor{\bsnm{{Bai}}, \binits{X.-N.}},
\bauthor{\bsnm{{Goodman}}, \binits{J.}}:
\bjtitle{\apj}
\bvolume{874}(\bissue{1}),
\bfpage{90}
(\byear{2019}).
\arxivurl{1810.12330}.
doi:\doiurl{10.3847/1538-4357/ab06fd}
\end{barticle}
\endbibitem

\bibitem[\protect\citeauthoryear{{Wang} et~al.}{2013}]{Wang2013}
\begin{barticle}
\bauthor{\bsnm{{Wang}}, \binits{Q.D.}},
\bauthor{\bsnm{{Nowak}}, \binits{M.A.}},
\bauthor{\bsnm{{Markoff}}, \binits{S.B.}},
\bauthor{\bsnm{{Baganoff}}, \binits{F.K.}},
\bauthor{\bsnm{{Nayakshin}}, \binits{S.}},
\bauthor{\bsnm{{Yuan}}, \binits{F.}},
\bauthor{\bsnm{{Cuadra}}, \binits{J.}},
\bauthor{\bsnm{{Davis}}, \binits{J.}},
\bauthor{\bsnm{{Dexter}}, \binits{J.}},
\bauthor{\bsnm{{Fabian}}, \binits{A.C.}},
\bauthor{\bsnm{{Grosso}}, \binits{N.}},
\bauthor{\bsnm{{Haggard}}, \binits{D.}},
\bauthor{\bsnm{{Houck}}, \binits{J.}},
\bauthor{\bsnm{{Ji}}, \binits{L.}},
\bauthor{\bsnm{{Li}}, \binits{Z.}},
\bauthor{\bsnm{{Neilsen}}, \binits{J.}},
\bauthor{\bsnm{{Porquet}}, \binits{D.}},
\bauthor{\bsnm{{Ripple}}, \binits{F.}},
\bauthor{\bsnm{{Shcherbakov}}, \binits{R.V.}}:
\bjtitle{Science}
\bvolume{341}(\bissue{6149}),
\bfpage{981}
(\byear{2013}).
\arxivurl{1307.5845}.
doi:\doiurl{10.1126/science.1240755}
\end{barticle}
\endbibitem

\bibitem[\protect\citeauthoryear{{Wu} et~al.}{2017}]{Wu2017}
\begin{barticle}
\bauthor{\bsnm{{Wu}}, \binits{M.-C.}},
\bauthor{\bsnm{{Bu}}, \binits{D.-F.}},
\bauthor{\bsnm{{Gan}}, \binits{Z.-M.}},
\bauthor{\bsnm{{Yuan}}, \binits{Y.-F.}}:
\bjtitle{\aap}
\bvolume{608},
\bfpage{114}
(\byear{2017}).
doi:\doiurl{10.1051/0004-6361/201730803}
\end{barticle}
\endbibitem

\bibitem[\protect\citeauthoryear{{Xue} and {Wang}}{2005}]{Xue2005}
\begin{barticle}
\bauthor{\bsnm{{Xue}}, \binits{L.}},
\bauthor{\bsnm{{Wang}}, \binits{J.}}:
\bjtitle{\apj}
\bvolume{623},
\bfpage{372}
(\byear{2005}).
doi:\doiurl{10.1086/428338}
\end{barticle}
\endbibitem

\bibitem[\protect\citeauthoryear{{Yan} and {Yu}}{2017}]{Yan2017}
\begin{barticle}
\bauthor{\bsnm{{Yan}}, \binits{Z.}},
\bauthor{\bsnm{{Yu}}, \binits{W.}}:
\bjtitle{\mnras}
\bvolume{470}(\bissue{4}),
\bfpage{4298}
(\byear{2017}).
\arxivurl{1706.06472}.
doi:\doiurl{10.1093/mnras/stx1562}
\end{barticle}
\endbibitem

\bibitem[\protect\citeauthoryear{{Yuan} and {Narayan}}{2014}]{Yuan2014}
\begin{barticle}
\bauthor{\bsnm{{Yuan}}, \binits{F.}},
\bauthor{\bsnm{{Narayan}}, \binits{R.}}:
\bjtitle{\araa}
\bvolume{52},
\bfpage{529}
(\byear{2014}).
\arxivurl{1401.0586}.
doi:\doiurl{10.1146/annurev-astro-082812-141003}
\end{barticle}
\endbibitem

\bibitem[\protect\citeauthoryear{{Yuan} et~al.}{2012}]{Yuan2012a}
\begin{barticle}
\bauthor{\bsnm{{Yuan}}, \binits{F.}},
\bauthor{\bsnm{{Bu}}, \binits{D.}},
\bauthor{\bsnm{{Wu}}, \binits{M.}}:
\bjtitle{\apj}
\bvolume{761},
\bfpage{130}
(\byear{2012}).
\arxivurl{1206.4173}.
doi:\doiurl{10.1088/0004-637X/761/2/130}
\end{barticle}
\endbibitem

\bibitem[\protect\citeauthoryear{{Yuan} et~al.}{2002}]{Yuan2002}
\begin{barticle}
\bauthor{\bsnm{{Yuan}}, \binits{F.}},
\bauthor{\bsnm{{Markoff}}, \binits{S.}},
\bauthor{\bsnm{{Falcke}}, \binits{H.}}:
\bjtitle{\aap}
\bvolume{383},
\bfpage{854}
(\byear{2002}).
\arxivurl{astro-ph/0112464}.
doi:\doiurl{10.1051/0004-6361:20011709}
\end{barticle}
\endbibitem

\bibitem[\protect\citeauthoryear{{Yuan} et~al.}{2012}]{Yuan2012}
\begin{barticle}
\bauthor{\bsnm{{Yuan}}, \binits{F.}},
\bauthor{\bsnm{{Wu}}, \binits{M.}},
\bauthor{\bsnm{{Bu}}, \binits{D.}}:
\bjtitle{\apj}
\bvolume{761},
\bfpage{129}
(\byear{2012}).
\arxivurl{1206.4157}.
doi:\doiurl{10.1088/0004-637X/761/2/129}
\end{barticle}
\endbibitem

\bibitem[\protect\citeauthoryear{{Yuan} et~al.}{2015}]{Yuan2015}
\begin{barticle}
\bauthor{\bsnm{{Yuan}}, \binits{F.}},
\bauthor{\bsnm{{Gan}}, \binits{Z.}},
\bauthor{\bsnm{{Narayan}}, \binits{R.}},
\bauthor{\bsnm{{Sadowski}}, \binits{A.}},
\bauthor{\bsnm{{Bu}}, \binits{D.}},
\bauthor{\bsnm{{Bai}}, \binits{X.-N.}}:
\bjtitle{\apj}
\bvolume{804},
\bfpage{101}
(\byear{2015}).
\arxivurl{1501.01197}.
doi:\doiurl{10.1088/0004-637X/804/2/101}
\end{barticle}
\endbibitem

\bibitem[\protect\citeauthoryear{{Yuan} et~al.}{2018}]{Yuan2018}
\begin{barticle}
\bauthor{\bsnm{{Yuan}}, \binits{F.}},
\bauthor{\bsnm{{Yoon}}, \binits{D.}},
\bauthor{\bsnm{{Li}}, \binits{Y.-P.}},
\bauthor{\bsnm{{Gan}}, \binits{Z.-M.}},
\bauthor{\bsnm{{Ho}}, \binits{L.C.}},
\bauthor{\bsnm{{Guo}}, \binits{F.}}:
\bjtitle{\apj}
\bvolume{857}(\bissue{2}),
\bfpage{121}
(\byear{2018}).
\arxivurl{1712.04964}.
doi:\doiurl{10.3847/1538-4357/aab8f8}
\end{barticle}
\endbibitem

\bibitem[\protect\citeauthoryear{{Zahra Zeraatgari}
  et~al.}{2018}]{Zeraatgari2018}
\begin{barticle}
\bauthor{\bsnm{{Zahra Zeraatgari}}, \binits{F.}},
\bauthor{\bsnm{{Mosallanezhad}}, \binits{A.}},
\bauthor{\bsnm{{Abbassi}}, \binits{S.}},
\bauthor{\bsnm{{Yuan}}, \binits{Y.-F.}}:
\bjtitle{\apj}
\bvolume{852}(\bissue{2}),
\bfpage{124}
(\byear{2018}).
\arxivurl{1712.03078}.
doi:\doiurl{10.3847/1538-4357/aa9ffd}
\end{barticle}
\endbibitem

\bibitem[\protect\citeauthoryear{{Zanni} et~al.}{2007}]{Zanni2007}
\begin{barticle}
\bauthor{\bsnm{{Zanni}}, \binits{C.}},
\bauthor{\bsnm{{Ferrari}}, \binits{A.}},
\bauthor{\bsnm{{Rosner}}, \binits{R.}},
\bauthor{\bsnm{{Bodo}}, \binits{G.}},
\bauthor{\bsnm{{Massaglia}}, \binits{S.}}:
\bjtitle{\aap}
\bvolume{469}(\bissue{3}),
\bfpage{811}
(\byear{2007}).
\arxivurl{astro-ph/0703064}.
doi:\doiurl{10.1051/0004-6361:20066400}
\end{barticle}
\endbibitem

\bibitem[\protect\citeauthoryear{{Zhang} and {Dai}}{2008}]{Zhang2008}
\begin{barticle}
\bauthor{\bsnm{{Zhang}}, \binits{D.}},
\bauthor{\bsnm{{Dai}}, \binits{Z.G.}}:
\bjtitle{\mnras}
\bvolume{388}(\bissue{3}),
\bfpage{1409}
(\byear{2008}).
\arxivurl{0805.3254}.
doi:\doiurl{10.1111/j.1365-2966.2008.13483.x}
\end{barticle}
\endbibitem

\bibitem[\protect\citeauthoryear{{Zhang} et~al.}{2012}]{Zhang2012}
\begin{barticle}
\bauthor{\bsnm{{Zhang}}, \binits{S.-N.}},
\bauthor{\bsnm{{Liao}}, \binits{J.}},
\bauthor{\bsnm{{Yao}}, \binits{Y.}}:
\bjtitle{\mnras}
\bvolume{421}(\bissue{4}),
\bfpage{3550}
(\byear{2012}).
\arxivurl{1201.3451}.
doi:\doiurl{10.1111/j.1365-2966.2012.20579.x}
\end{barticle}
\endbibitem

\end{thebibliography}

\end{document}